%% file: 0-main.tex
\documentclass[lettersize,journal]{IEEEtran}
\usepackage{cite}
\usepackage{booktabs}
\usepackage{hyperref}
\usepackage{amsmath,amsfonts,amssymb}
\usepackage{amsthm} 
\usepackage{array}
\usepackage[caption=false,font=normalsize]{subfig}
\usepackage{textcomp}
\usepackage{stfloats}
\usepackage{url}
\usepackage{verbatim}
\usepackage{graphicx}
\usepackage{caption}
\usepackage{pifont}
\usepackage[noend]{algpseudocode}
\usepackage{graphicx}
\usepackage{color}
\usepackage{algpseudocode}  
\usepackage[linesnumbered, ruled, vlined, shortend]{algorithm2e}
\usepackage{verbatim}
\usepackage[graphicx]{realboxes}
\usepackage{rotating}
\usepackage{array}
\usepackage{threeparttable}
\usepackage{makecell}
\usepackage{soul}
\usepackage{tabularx}
\usepackage{xcolor}
\usepackage[most]{tcolorbox}
\tcbuselibrary{breakable}
\usepackage{enumitem}
\usepackage{multirow}
\newcommand{\framework}{CoDeSEG}
\newtheorem{definition}{Definition}
\usepackage{mathrsfs}
\usepackage{xcolor}
\usepackage{colortbl}
  
\begin{document}

\title{Comprehensive, Efficient Large-Scale Community Detection via Structural Entropy Game}

\author{Pu Li, Yantuan Xian, Hao Peng, Huafeng Li, Zhengtao Yu, Yan Xiang, Philip S. Yu,~\IEEEmembership{Fellow,~IEEE} 
\IEEEcompsocitemizethanks{
\IEEEcompsocthanksitem Pu Li, Yantuan Xian, Huafeng Li, Zhengtao Yu, and Yan Xiang are with the School of Information Engineering and Automation, Kunming University of Science and Technology, and Yunnan Key Laboratory of Artificial Intelligence, Kunming 650500, China. E-mail: lip@stu.kust.edu.cn, \{xianyt, yuzt, yanx\}@kust.edu.cn,  hfchina99@163.com.
\IEEEcompsocthanksitem Hao Peng is with the School of Cyber Science and Technology, Beihang University, Beijing 100191, China. E-mail: penghao@buaa.edu.cn.
\IEEEcompsocthanksitem  Philip S. Yu is with the Department of Computer Science, University of Illinois at Chicago, Chicago, IL 60607, USA. E-mail: psyu@uic.edu.
}
\thanks{Corresponding author: Yantuan Xian}
}
\markboth{Journal of XXX}%
{Li \MakeLowercase{\textit{et al.}}: Community Detection in Large-Scale Complex Networks via Structural Entropy Game}


\maketitle

\begin{abstract}
Community detection is a critical task in graph theory, social network analysis, and bioinformatics, where communities are defined as clusters of densely interconnected nodes. 
However, detecting communities in large-scale networks with millions of nodes and billions of edges remains challenging due to the inefficiency and unreliability of existing methods. 
Moreover, many existing methods are limited to specific types of graph structures (such as unweighted or undirected graphs) or are designed solely for detecting static communities, reducing their broader applicability.
To address these issues, we propose a novel heuristic community detection algorithm, termed \framework, which identifies communities by minimizing the network's two-dimensional (2D) structural entropy within a potential game framework. 
In the game, nodes decide to stay in the current community or move to another based on a strategy that maximizes the 2D structural entropy utility function.
Additionally, we introduce a structural entropy-based node overlapping heuristic for detecting overlapping communities, with a near-linear time complexity.
Furthermore, we design a cascading influence propagation-based adaptive community update strategy, which dynamically identifies and processes nodes whose community affiliations may change during graph evolution, thereby effectively extending \framework~to dynamic community detection scenarios.
Experimental results on fourteen large-scale networks demonstrate that \framework~ achieves state-of-the-art performance across three community detection tasks (overlapping, non-overlapping, dynamic), while also delivering substantial improvements in detection efficiency.

\end{abstract}


\section{Introduction}
\label{sec:introduction}
\input{1_introduction}

\section{Preliminaries}
\label{sec:problem_definition}
\input{3_problem_definition}

\section{Methodology}
\label{sec:methodology}
\input{4_methodology}

\section{Experimental Setup}
\label{sec:Setupp}
\input{5_experimental_setup}

\section{Experiments}
\label{sec:experiments}
\input{6_experiments}

\section{\textcolor{black}{Related Work}}
\label{sec:related_work}
\input{2_related_work}

\section{Conclusion}
\label{sec:conclusion}
\input{7_conclusion}

\bibliographystyle{IEEEtran}
\bibliography{8_references}
\clearpage

\input{Appendix}

\end{document}

%% file: 1_introduction.tex
\begin{figure}[t]
    \centering
    \includegraphics[width=1\linewidth, trim={0.1cm 0.2cm 1.8cm 0.1cm}, clip]{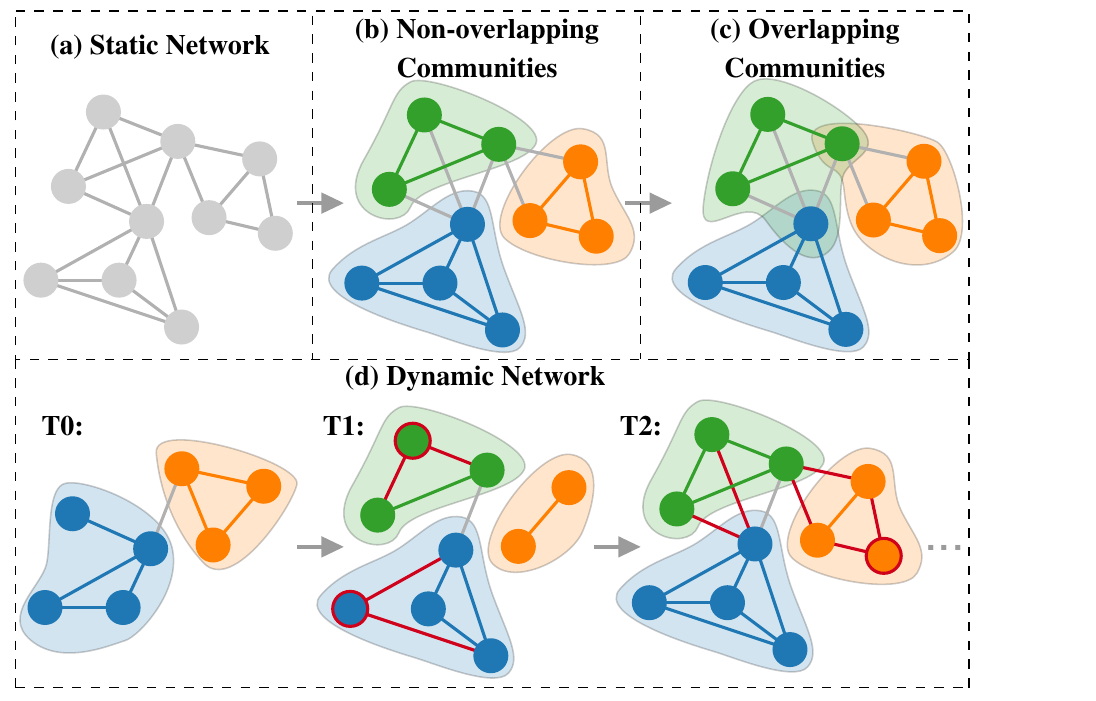}
    \caption{Illustration of different community structures in a network.} \vspace{-4mm}
    \label{Figtask}
\end{figure}

\IEEEPARstart{C}{ommunity} refers to a set of closely related nodes within a network, also known as a cluster or module in literature~\cite{Fortunato2010Community,Fortunato2022}. 
Community detection is a task that reveals fundamental structural information within real-world networks, providing valuable insights by identifying tightly knit subgroups. 
In drug discovery, for instance, detecting protein functional groups facilitates the identification of novel, valuable proteins~\cite{ma2019comparative}. 
In social event detection, analyzing message groups within social streams helps to understand the development trends of events and analyze public sentiment~\cite{cao2024hierarchical}. Community detection also plays a role in recent retrieval-augmented generation (RAG) applications, like GraphRAG \cite{edge2024localglobalgraphrag}. Furthermore, community detection has extensive applications across various domains, including recommender systems~\cite{basuchowdhuri2014analysis}, medicine~\cite{barabasi2011network}, biomedical research~\cite{manipur2021community}, social networks~\cite{fortunato2016community,pengunsupervised}, and more.
As illustrated in Figure~\ref{Figtask}(b), most early research on community detection has focused on disjoint clusters, where each node belongs to a single community, and there is no overlap between communities \cite{Fortunato2010Community, traag2019louvain, blondel2008fast, Raghavan2007Near,caomulti,yang2024incremental}. 
However, nodes often participate in multiple communities in many real-world applications (as depicted in Figure~\ref{Figtask}(c)), sparking a growing interest in detecting overlapping communities \cite{palla2005uncovering, xie2013overlapping,kelley2009existence}. 
Overlapping community detection typically entails higher computational costs and time overhead than disjoint community detection.
Over the past two decades, numerous algorithms for overlapping community detection have been proposed, including those based on modularity \cite{cherifi2019community}, label propagation~\cite{lu2018lpanni,xie2011slpa}, seed expansion~\cite{whang2013overlapping, asmi2022greedy}, non-negative matrix factorization~\cite{yang2013overlapping}, and spectral clustering~\cite{van2019scalable}.
However, existing overlapping community detection methods~\cite{chen2010game, Gopalan2013Efficient, yang2013overlapping} struggle to scale to networks with millions of nodes and billions of edges, often requiring days or longer to produce results.

Many well-established and widely-used methods for large-scale networks are typically limited to specific types of graphs, such as unweighted or undirected graphs, thereby restricting their applicability. 
For instance, algorithms like Bigclam~\cite{yang2013overlapping} and SLPA~\cite{xie2011slpa}  detect overlapping communities in unweighted and undirected graphs. Methods like Louvain~\cite{blondel2008fast}, Leiden~\cite{traag2019louvain}, and LPA~\cite{Raghavan2007Near} focus on detecting non-overlapping communities in undirected graphs. 
Thus, detecting overlapping communities in weighted, directed, large-scale networks remains a significant challenge.

Many real-world networks, particularly online social networks such as Facebook and Twitter, are inherently dynamic, with their network structures continually evolving over time.
Correspondingly, the community structures within these networks also exhibit significant dynamism (as shown in Figure~\ref{Figtask}(d)). 
Applying static community detection algorithms directly to network snapshots at different time points fails to fully utilize historical community information, and often results in substantially reduced computational efficiency as the network scale increases~\cite{zhuang2019dynamo}.
Contemporary incremental community detection methods attempt to address these challenges by updating community structures based on predefined heuristic rules~\cite{zhuang2019dynamo,seifikar2020c,nguyen2011adaptive,folino2013evolutionary} (e.g., node or edge insertions and deletions) or by extracting subgraphs affected by localized structural changes for partial re-optimization~\cite{sun2022dynamic,anuar2025bird,cordeiro2016dynamic}. 
Nevertheless, these approaches often face challenges with computational efficiency in large-scale dense networks and struggle to fully identify all affected regions of the graph. 
As the network evolves, they also struggle to maintain a proper trade-off between detection accuracy and computational efficiency\cite{anuar2025bird}.

To tackle these challenges, we propose a novel algorithm named \framework~(\textbf{Co}mmunity \textbf{De}tection via \textbf{S}tructural \textbf{E}ntropy \textbf{G}ame) in large-scale complex networks.
The proposed algorithm follows the game-theoretic inspired community formation game framework \cite{chen2010game}. 
In the game, we quantify the rewards and required costs associated with nodes joining or leaving a community through the changes in structural entropy.
The Nash equilibrium of the game directly corresponds to the network's community structure, with each node's community memberships at equilibrium serving as the output of the community detection algorithm. 
Specifically, we \textbf{first} define the potential function as the 2-dimensional structural entropy (2D SE)~\cite{li2016structural} of the network and derive an efficient node utility function computable in approximately constant time. 
By applying the node utility to the community formation game, we detect communities in large-scale networks efficiently.
\textbf{Second}, we present a structural entropy-based node overlap heuristic function to detect overlapping communities, which can leverage the intermediate results of the community formation game to speed up the algorithm. 
\textbf{Third},  we propose a cascading influence propagation-based adaptive community update strategy to extend our algorithm to dynamic network scenarios. 
This strategy comprises two core mechanisms: 
(1) hierarchical affected node identification, which detects directly affected nodes triggered by network changes (e.g., node/edge additions/deletions) and subsequently identifies indirectly affected nodes through cascading influence during node gaming;   
(2) adaptive node removal, which differentially evaluates the stability of nodes' community affiliations based on node hierarchy (direct/indirect influence), and removes stable nodes.
This strategy guides perturbed network regions to converge toward stable community structures during the multi-round game.
Notably, the proposed algorithm is applicable to diverse network types, including unweighted, weighted, undirected, and directed graphs, generating stable and reliable community structures within a unified paradigm. 
The algorithm's simplicity also facilitates straightforward parallelization, further enhancing its efficiency by allowing node strategies to be computed concurrently. 
Experiments on nine real-world networks and five synthetic networks demonstrate \framework's consistent superiority across three community detection tasks (overlapping, non-overlapping, dynamic), outperforming all baselines in performance while incurring significantly lower time overhead than the second-fastest baseline.
The codes of~\framework~and baselines, along with datasets, are publicly available on GitHub\footnote{\url{https://github.com/Kust-lp/CoDeSEG}}.

In summary, the contributions of this paper are as follows:
\begin{itemize}[left=0pt]
    \item We propose a novel heuristic algorithm for community detection in large-scale networks, termed ~\framework. 
    This algorithm introduces 2D SE to define the potential function of the community formation game and derives a node utility function with nearly constant time complexity.
    To the best of our knowledge, \framework~is the fastest known algorithm for large-scale network community detection.
    \item We design a two-stage algorithm for detecting overlapping communities in diverse graphs. 
    \framework~identifies non-overlapping communities through the proposed community formation game and subsequently detects overlapping communities rapidly using a node overlap heuristic function based on structural entropy.
    \item We propose a cascading influence propagation-based adaptive community update strategy that integrates hierarchical influential node identification with an adaptive node removal mechanism, thereby extending \framework~to dynamic network environments.
    This strategy comprehensively identifies all potentially affected nodes and minimizes computational overhead, thereby achieving a better trade-off between detection efficiency and quality.
    \item Experimental results on fourteen large-scale networks and three detection tasks (i.e., overlapping, non-overlapping, and dynamic detection) demonstrate that the \framework~ algorithm consistently outperforms state-of-the-art community detection methods in terms of ONMI, NMI, and F1-score, while also achieving substantial reductions in detection time. 
    Notably, compared to the fastest overlapping detection baseline, \framework~ achieves an average speedups of 33 times.
\end{itemize}

This paper is an extended version of our conference paper~\cite{xian2025community}, published at The Web Conference 2025. 
The journal version introduces the following major enhancements:
(1)  We propose a novel cascading influence propagation-based adaptive community update strategy, which enables the original framework to adapt to dynamic networks while preserving theoretical consistency (Section \ref{Secdynamic}).
(2) We conduct experiments on two large-scale real-world dynamic network datasets and provide detailed empirical analyses that comprehensively validate the effectiveness and stability of the proposed method in dynamic community detection tasks (Sections \ref{DCDE}, \ref{Effic} and \ref{SDN}).
In addition, we perform non-overlapping community detection experiments on five additional large-scale synthetic networks (Section \ref{NOCD});
(3) We present a more comprehensive discussion of related work, particularly in dynamic community detection and structural entropy (Section \ref{sec:related_work}).

%% file: 3_problem_definition.tex
In this section, we summarize the concepts and definitions related to the background of our work, including community detection, community formation games, and structural entropy.

\subsection{\textcolor{black}{Community Detection}}
Community detection aims to identify groups of nodes within a network that exhibit high internal cohesion and low external coupling.
Based on the temporal characteristics of the network, which can be categorized into static community detection and dynamic community detection.
\begin{definition}[\textcolor{black}{Static Community Detection}]
     Given a network \( G = (\mathcal{V}, \mathcal{E}) \), where \( \mathcal{V} \) is the set of nodes (vertices), \( \mathcal{E} \) is the set of edges (links), community detection algorithms find a set of communities \( \mathcal{P} = \{\mathcal{C}_1, \mathcal{C}_2, \dots, \mathcal{C}_k\} \), where each \( \mathcal{C}_i \subseteq \mathcal{V} \) is a network community. 
     In the \textbf{overlapping community detection}, nodes \( x \in \mathcal{V} \) can belong to more than one community.
\end{definition}

\begin{definition}[\textcolor{black}{Dynamic Community Detection}]
A dynamic network $G^d$ is defined as a temporal sequence of network snapshots: $G^d = \{ G_0, G_1, \dots, G_{t-1}, G_t \}$, where $G_t$ denotes the network snapshot at time $t$. 
The evolution between consecutive snapshots is denoted as: $G_t = G_{t-1} \cup \Delta G^{t}$.
Here, $ \Delta G^{t} = (\Delta \mathcal{V}^{t}, \Delta \mathcal{E}^{t})$ represents the incremental change from $G_{t-1}$ to $G_t$, $\Delta \mathcal{V}^{t}$ and $\Delta \mathcal{E}^{t}$ denote the sets of vertices and edges changed during the time interval $(t-1, t]$, respectively. 
A dynamic community detection algorithm incrementally updates the current network snapshot's communities based on the previous snapshot's communities. 
As defined by the function: $\mathcal{P}_t = f(G_{t-1}, \mathcal{P}_{t-1}, G_t)$, where $\mathcal{P}_{t-1}$ represents the community partition of network snapshot $G_{t-1}$.
\end{definition}

\subsection{Community Formation Game}
Chen et al. \cite{chen2010game} propose a game-theoretic-based community detection framework, named \textbf{community formation game}, that simulates the strategy selection and interactions of nodes within a network to identify community structures. 
In the game, each node \(x \in \mathcal{V}\) is treated as a rational participant (player), consistently choosing the best strategy (community) that maximizes utility function. 
When the game converges to a Nash equilibrium, it corresponds to the communities the algorithm detects. We present relevant definitions, as follows:

\begin{definition}[Strategy Profile]
A strategy profile is a combination of strategies chosen by all players in the game. If there are \( n \) players in the game, and each player \( i \) has a set of strategies \( S_i \), then a strategy profile \( s \) is a tuple \( \boldsymbol{s} = (s_1, s_2, \ldots, s_n) \), where \( s_i \in S_i \) is the strategy chosen by player \( i \).
\end{definition}

\begin{definition} [Utility Function]
A utility (payoff) function represents the benefit a player receives based on the chosen strategies. For a player \( i \), the utility function is denoted by \( u_i: S \to \mathbb{R} \), where \( S \) is the set of all possible strategy profiles. 
The function \( u_i(s) \) gives the payoff to player \( i \) when the strategy profile \( \boldsymbol{s} \) is played.
\end{definition}

\begin{definition}[Potential Game]
There exists a potential function \( \varphi: S \to \mathbb{R} \), for any player \( i \) and any two strategy profiles \( \boldsymbol{s} \) and \( \boldsymbol{s}^\prime \) differing only in the strategy of player \( i \), the change in the potential function equals the change of player \( i \)'s payoff:
\begin{equation}
    \varphi(\boldsymbol{s}^\prime) - \varphi(\boldsymbol{s}) = u_i(s^\prime) - u_i(s)
\end{equation}
where \( u_i \) is the utility function for player \( i \). 
Algorithms for learning in potential games, such as best response dynamics, are capable of converging to a Nash equilibrium. The Nash equilibrium refers to a stable state in multiplayer games in which no player can improve their outcome by unilaterally altering their strategy. 
\end{definition}

\subsection{\textcolor{black}{Structural Entropy}}
Structural entropy (SE) quantifies uncertainty and information content in complex networks, with lower values indicating more ordered structures and higher values reflecting greater disorder~\cite{li2016structural}. 
Communities within a network can be identified by minimizing its 2D SE. 
\begin{definition}[\textcolor{black}{2D SE}]
Suppose $\mathcal{P}=\{\mathcal{C}_1, \mathcal{C}_2, \dots, \mathcal{C}_k\}$ is a partition of the network $G$, the 2D SE is defined as:
\begin{equation}
    \begin{aligned}
        \mathcal{H}^{2}(\mathcal{P})  
        = - \sum_{c \in \mathcal{P} } \left( \frac{g_{c}}{v_\lambda} \log \frac{v_c}{v_\lambda} + \sum_{x \in c }  \frac{d_x}{v_\lambda} \log \frac{d_x}{v_c} \right),
    \end{aligned}
\end{equation}
where $d_x$ is the degree of node $x$, $v_\lambda$ is the network's volume.
$g_c$ and $v_c$ denote the volume and cut of $c$, respectively.
For a more detailed definition of SE, refer to \textbf{ Appendix B}.
\end{definition}

%% file: 4_methodology.tex
This section introduces the proposed algorithm, \framework.
Section~\ref{SecSEHeuristic} presents the structural entropy-based heuristic for the community formation game, followed by Section~\ref{SecComputeDeltaLeave}, which details key strategy computations.
The complete static community detection algorithm is outlined in Section~\ref{SecAlgorithm}, while Section~\ref{Secdynamic} introduces the Cascading Influence Propagation-based Adaptive (CIPA) community update strategy for dynamic community detection.
Furthermore, Section~\ref{SecTimeComplexity} analyzes the time and space complexity of \framework, and Section~\ref{SecParallel} describes its parallel implementation.

\subsection{Structural Entropy based Heuristic Function}
\label{SecSEHeuristic}
The proposed algorithm models community formation as a potential game, where the potential function is the network's 2D SE $\mathcal{H}^{2}(\mathcal{P})$. Each node selects the community that most reduces this entropy as its optimal strategy.
When the game converges to a Nash equilibrium, yielding communities with a minimized 2D SE.
Consider a node in $G$ adopting a strategy, such as altering its community membership, resulting in a new partition denoted by $\mathcal{P}^\prime$.
We define the \textbf{heuristic function} $\Delta$ as the change in the potential function:
\begin{equation}
    \Delta = \mathcal{H}^{2}(\mathcal{P}) - \mathcal{H}^{2}(\mathcal{P}^\prime).
\end{equation}

In the disjoint community formation game, each node aims to maximize the value of $\Delta$ by moving to the best adjacent community, resulting in a partition with reduced 2D structural entropy.
A node can choose from three strategies: \textbf{Stay}, \textbf{Leave and be alone}, and \textbf{Transfer to another community}. 

\textbf{Stay}. 
Node $x$ decides to stay in the current community, then the partition $\mathcal{P}$ remains unchanged, $\mathcal{P}^\prime=\mathcal{P}$. The value of heuristic function $\Delta_{\text{S}}$ is:
\begin{equation}
    \Delta_{\text{S}} = \mathcal{H}^{2}(\mathcal{P}) - \mathcal{H}^{2}(\mathcal{P}^\prime) = 0.
\end{equation}

\textbf{Leave and be alone}. 
Suppose the original partition is $\mathcal{P}=\{\mathcal{C}_1, \mathcal{C}_2, \dots, \mathcal{C}_k\}$ and when node $x$ leaves its community $\mathcal{C}_k$, resulting a new partition $\mathcal{P}^\prime=\{\mathcal{C}_1, \mathcal{C}_2, \dots, \mathcal{C}_k^\prime, \{ x\}\}$, where $\mathcal{C}_k = \mathcal{C}_k^\prime \cup \{x\}$. 
The value of heuristic function $\Delta_{\text{L}}(x, \mathcal{C}_k)$ is:
\begin{equation}
\label{EqDeltaLeave}
\begin{aligned}
    \Delta_{\text{L}}(x, \mathcal{C}_k) 
    =& \mathcal{H}^{2}(\mathcal{P}) - \mathcal{H}^{2}(\mathcal{P}^\prime) \\
    =& \mathcal{H}^{2}(\mathcal{C}_k) - \mathcal{H}^{2}(\mathcal{C}_k^\prime) - \mathcal{H}^{2}(\{x\}).
\end{aligned}
\end{equation}
The calculation details for Equation~(\ref{EqDeltaLeave}) are provided in Section~\ref{SecComputeDeltaLeave}. 
If community $\mathcal{C}_k$ is a singleton, then $\Delta_\text{L}(x, \mathcal{C}_k) = 0$.

\textbf{Transfer to another community}.
Suppose $x$ transfers from $\mathcal{C}_1$ to $\mathcal{C}_k$, the original partition is $\mathcal{P}=\{\mathcal{C}_1, \mathcal{C}_2, \dots, \mathcal{C}_k\}$ and the new partition is $\mathcal{P}^\prime=\{\mathcal{C}_1^\prime, \mathcal{C}_2, \dots, \mathcal{C}_k^\prime\}$, where $\mathcal{C}_1^\prime = \mathcal{C}_1 \backslash \{x\}$, $\mathcal{C}_k^\prime = \mathcal{C}_k \cup \{x\}$.
The transfer strategy can be seen as a composite transformation of two steps.
Firstly, node $x$ leaves $\mathcal{C}_1$ and does not join any community, which yields an intermediate $\mathcal{P}^m = \{ \mathcal{C}_1^\prime, \mathcal{C}_2, \dots, \mathcal{C}_k, \{x\}  \}$. 
Secondly, node $x$ join $\mathcal{C}_k$, resulting in new partition $\mathcal{P}^\prime$. 
We can easily figure out that the second step is an inverse transformation of the Leave and be alone strategy. 
Therefore, the value of the heuristic function $\Delta_{\text{T}}(x, \mathcal{C}_1,  \mathcal{C}_k)$ can be expressed as:
\begin{equation}
\label{EqDeltaTransfer}
\begin{aligned}
    \Delta_{\text{T}}&(x, \mathcal{C}_1, \mathcal{C}_k) = \mathcal{H}^{2}(\mathcal{P}) - \mathcal{H}^{2}(\mathcal{P}^\prime)\\
    &= (\mathcal{H}^{2}(\mathcal{P}) -\mathcal{H}^{2}(\mathcal{P}^m)) + (\mathcal{H}^{2}(\mathcal{P}^m) - \mathcal{H}^{2}(\mathcal{P}^\prime)) \\
    &= \Delta_{\text{L}}(x, \mathcal{C}_1) - \Delta_{\text{L}}(x, \mathcal{C}_k).
\end{aligned}
\end{equation}
Since a node $x$ may have several adjective target communities, we denote the best-transferring with max $\Delta_{\text{T}}$ as $\Delta_{\text{T}}(x, \mathcal{C}_{\text{best}})$.
\begin{figure}[b]
    \centering
    \includegraphics[width=1\linewidth, trim={0.1cm 0cm 3.3cm 0.1cm}, clip]{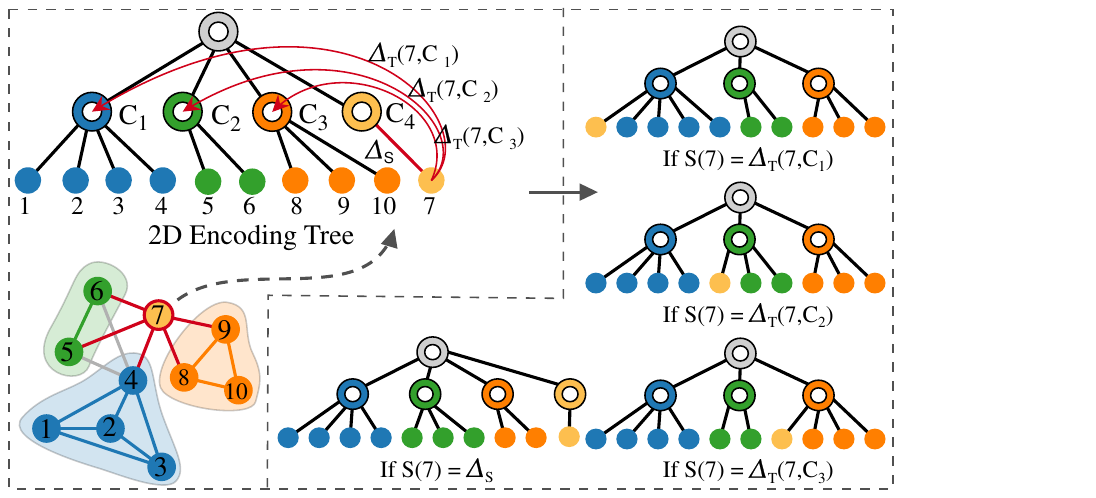}
    \caption{\textcolor{black}{A toy of the 2D Encoder Tree and potential game.}}
    \label{encoder}
\end{figure}
In our community formation game, a node will only join a new community if it decreases the network's 2D structural entropy.
Consequently, a node will prefer to stay in its current community unless another community offers a further reduction in 2D structural entropy. 
Thus, the Leave and be alone strategy is optional. 
As shown in Figure~\ref{encoder},  node \( x \) selects its movement strategy according to the following formula:
\begin{equation}
\label{EqStrategy}
    S(x) = \max (\Delta_\text{S}, \Delta_{\text{T}}(x, \mathcal{C}_{\text{best}})).
\end{equation}
After identifying disjoint communities, we propose using a SE heuristic to determine whether a node \( x \) should overlap between communities.

\textbf{Overlap nodes}.
Suppose partition $\mathcal{P} = \{\mathcal{C}_1, \mathcal{C}_2, \dots, \mathcal{C}_k\}$, for a node $ x \notin \mathcal{C}_k $, exists
$y \in \text{Neighbors}(x), \ y \in \mathcal{C}_k$. 
If we copy (overlap) \( x \) to community $\mathcal{C}_k$, we create a new overlapping partition $\mathcal{P}^\prime = \{\mathcal{C}_1, \mathcal{C}_2, \dots, \mathcal{C}_k^\prime\}$, where $\mathcal{C}_k^\prime = \mathcal{C}_k \cup \{x\}$.
Since the copy action does not affect the origin community, we define the overlapping heuristic function as follows:
\begin{equation}
\label{EqOverlap}
\begin{aligned}
    \Delta_{\text{O}}(x, \mathcal{C}_k) = \mathcal{H}^{2}(\mathcal{P}) - \mathcal{H}^{2}(\mathcal{P}^\prime) 
        = - \Delta_L(x, \mathcal{C}_k^\prime). \\
\end{aligned}
\end{equation}
If \( \Delta_\text{O}(x, \mathcal{C}_k) > 0 \), it means the overlap action reduces the partition's structural entropy. 
However, this criterion may allow excessive node overlapping. 
To address this, we propose using the average value of $\Delta_\text{L}(x_i, \mathcal{C}_k)$ for nodes in the community $\mathcal{C}_k$ as a threshold.
Nodes can only overlap with $\mathcal{C}_k$ if their $\Delta_\text{L}(x, \mathcal{C}_k)$ exceeds this threshold.

\begin{figure*}[th]
    \centering
    \includegraphics[width=1\linewidth]{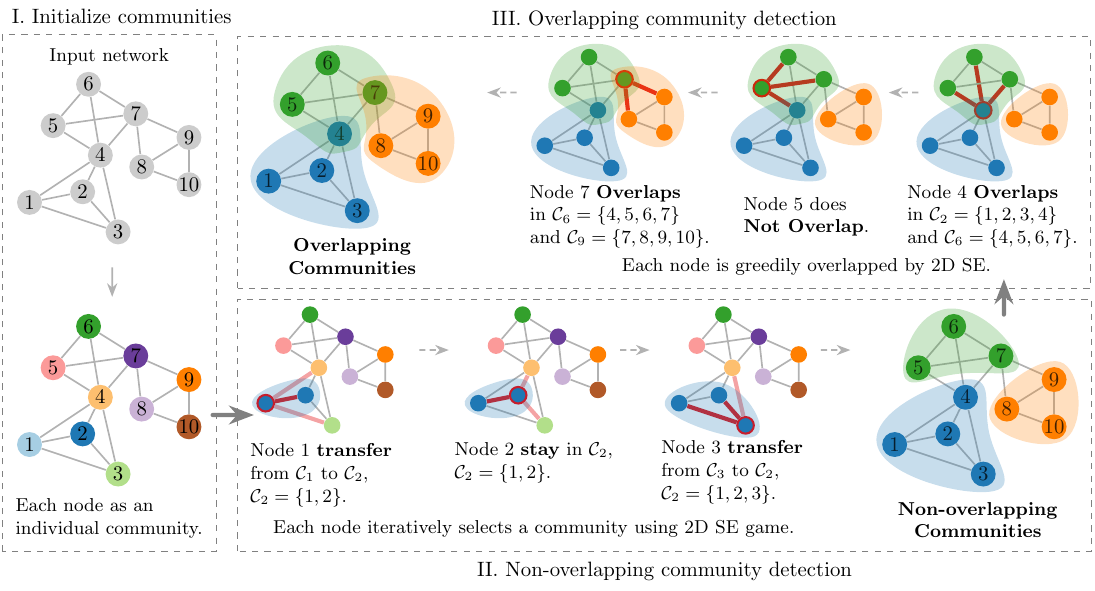}
    \vspace{-4mm}
    \caption{Overview of the proposed \framework~ algorithm.\vspace{-4mm}}
    \label{FigOverview}
\end{figure*}

\subsection{Efficient computation of $\Delta_{\text{L}}(x, \mathcal{C}_k)$}
\label{SecComputeDeltaLeave}
The formulas for the strategies above highlight that efficiently completing the community formation game depends on quickly computing the node leave strategy. By deriving Equation (\ref{EqDeltaLeave}), we obtain an efficient formula for computing $\Delta_{\text{L}}(x, \mathcal{C}_k)$:
\begin{equation}
\label{EqEfficentDeltaLeave}
\begin{aligned}
    &\Delta_{\text{L}}(x, \mathcal{C}_k) = \mathcal{H}^{2}(\mathcal{P}) - \mathcal{H}^{2}(\mathcal{P}^\prime) \\
    &= \frac{g_{\mathscr{c}_k^\prime}}{v_\lambda} \log \frac{v_{\mathscr{c}_k^\prime}}{v_\lambda}  - \frac{g_{\mathscr{c}_k}}{v_\lambda} \log \frac{v_{\mathscr{c}_k}}{v_\lambda} 
        + \frac{d_x}{v_\lambda} \log \frac{v_{\mathscr{c}_k}}{v_\lambda}
        + \frac{v_{\mathscr{c}_k^\prime}}{v_\lambda}  \log \frac{v_{\mathscr{c}_k}}{v_{\mathscr{c}_k^\prime}},
\end{aligned}
\end{equation}
where \( v_{\mathscr{c}_k^\prime} \) represents the volume of $\mathcal{C}_k^\prime$, and \( g_{\mathscr{c}_k^\prime} \) denotes the sum of the degrees (weights) of the cut edges of $\mathcal{C}_k^\prime$. The detailed derivation is provided in \textbf{Appendix C}.

By caching all community volumes \(\{ v_{c_1}, v_{c_2}, \dots, v_{c_k} \}\) and cut edge summations \(\{ g_{c_1}, g_{c_2}, \dots, g_{c_k} \}\), computing \(v_{c_k^\prime}\) and \(g_{c_k^\prime}\) becomes straightforward, allowing us to calculate \(\Delta_{\text{L}}(x, \mathcal{C}_k)\) in constant time complexity, \(O(1)\). 
For an undirected graph, \(v_{c_k^\prime}\) and \(g_{c_k^\prime}\) can be computed using the following equations:
\begin{equation}
\label{EqUpdateVolCut}
        v_{c_k^\prime} =v_{c_k} - d_x, \quad
        g_{c_k^\prime} = g_{c_k} + 2 d_x^{\text{in}} - d_x,
\end{equation}
where $d_x^{\text{in}}$ denotes the sum of edge weights between node $x$ and its neighbor nodes within community $\mathcal{C}_k$. 
The proposed strategies can be easily adapted for directed networks by separately considering the in-degree and out-degree of node \( x \) relative to community \(\mathcal{C}_k\). Consequently, Equation~(\ref{EqUpdateVolCut}) is updated accordingly:
\begin{equation}
        v_{c_k^\prime} = v_{c_k} - d_x, \quad
        g_{c_k^\prime} = g_{c_k} +  d_x^{\text{in}} +d_x^{\text{out}} - d_x .
\end{equation}
Once a node $x$ is transferred to a new community, we will update the statistics of the target and source communities using the following formulas,
\begin{equation}
\label{EqUpdateTarget}
\begin{array}{ll}
v_{c_t} \gets v_{c_t} + d_x, & v_{c_k} \gets v_{c_k^\prime},\\
g_{c_t} \gets g_{c_t} - 2d_x^{\text{in}} + d_x, & g_{c_k} \gets g_{c_k^\prime}.\\
\end{array}
\end{equation}
For directed graphs, we can easily derive similar formulas for updating community statistics.

\begin{algorithm}[t]

    \SetAlgoShortEnd
    \caption{Non-overlapping Community Detection.\label{AlgNonOverlapDetect}}
    \let\oldnl\nl
    \newcommand{\nonl}{\renewcommand{\nl}{\let\nl\oldnl}}

    \DontPrintSemicolon
    \KwIn{Graph $ G = (\mathcal{V}, \mathcal{E})$; Termination threshold $\tau_n$.}
    \KwOut{Non-overlapping communities $\mathcal{P}$ of $G$.}
    
    $\mathcal{P} \gets $ Each node as an individual community.\;
    Initialize community volumes \( \{ v_{c_1}, v_{c_2}, \dots, v_{c_n} \}\).\;
    Initialize cut edge summations \( \{ g_{c_1}, g_{c_2}, \dots, g_{c_n} \}\).\;

    \While{true}{
        $\Delta_{\text{sum}} \gets 0, M \gets 0$\;

        \For{node $x \in \mathcal{V}$}{
            $\mathcal{C}_x \gets $ Community contains $x$ \; 
            $t_c \gets $ Index of community $\mathcal{C}_x$ \;
            $t \gets t_c$ \Comment{$t_c$ means stay in $\mathcal{C}_x$ } \;

             $\Delta_{\text{L}}(x, \mathcal{C}_x) \gets $ Eq.\ref{EqEfficentDeltaLeave} \;

            $\Delta_{\max} \gets \Delta_{\text{L}}(x, \mathcal{C}_x)$\;
    
            \For{$k$-th adjacent community $\mathcal{C}_k$ of node $x$}{
                $\Delta_{\text{L}}(x, \mathcal{C}_k) \gets$ Eq.\ref{EqEfficentDeltaLeave} \;
                $\Delta_{\text{T}}(x, \mathcal{C}_x, \mathcal{C}_k) \gets $Eq.\ref{EqDeltaTransfer} \;

                \If{$\Delta_{\text{T}}(x, \mathcal{C}_x, \mathcal{C}_k) > \Delta_{\max}$}{
                    $\Delta_{\max} \gets \Delta_{\text{T}}(x, \mathcal{C}_x, \mathcal{C}_k)$\;
                    $t \gets k$ \;
                }
            }
    
            \If{$t \neq t_c$}{
                Transfer $x$ from $\mathcal{C}_x$ to $\mathcal{C}_t$ \;
                Update statistics of $\mathcal{C}_x$, $\mathcal{C}_t$ by Eq.\ref{EqUpdateTarget} \;
                $M \gets M+1$. \;
            }

            $\Delta_{\text{sum}} \gets \Delta_{\text{sum}} + \Delta_{\max}$ \;
        }
        
        \If{$M = 0$ or Eq.~\ref{eq:termination} is satisfied}{
            Break \;
        }
    }    
    \Return{$\mathcal{P}$}
    
\end{algorithm}
\subsection{Static Community Detection Algorithm}
\label{SecAlgorithm}

After developing effective methodologies for community formation games, we introduce a new two-stage algorithm for overlapping community detection. Figure~\ref{FigOverview} provides an overview of this algorithm.
In the non-overlapping phase, each node \( x \) sequentially implements the best strategy from Equation (\ref{EqStrategy}) until all nodes are assigned to their communities. 
In the overlapping phase, nodes \( x \) can overlap multiple communities if the overlap action meets the specified threshold.

\textbf{Non-overlapping Community Detection.}\label{SecNonOverlappingDetection} 
We propose a non-overlapping community detection algorithm designed to minimize 2D SE using a potential game, where the optimal strategy is computed by Equation (\ref{EqStrategy}). 
Once the game reaches a Nash equilibrium, the communities stabilize and no longer change. 
As shown in Algorithm~\ref{AlgNonOverlapDetect}, we initialize each node as an individual cluster and compute their volumes, setting the cut edges' summations as the node degrees (lines 1-3). 
In a directed network, the volumes of these communities are node in-degrees, and the summations of cut edges are node out-degrees.
The heart of our algorithm lies in an iterative loop of community formation games. 
We evaluate every node \( x \) in each iteration and determine the optimal strategy for \( x \) to significantly minimize the 2D SE of the graph (lines 7-17). 
To get the best strategy for a node, we first compute the heuristic \( \Delta_{\text{L}}(x, \mathcal{C}_x) \) that provides a measure of the impact when node \( x \) leaves its current community. 
The maximum heuristic function value \( \Delta_{\max} \) is initially set to \( \Delta_{\text{L}}(x, \mathcal{C}_x) \), and the target community index \( t \) is set to \(t_c\) indicating stay in the current community (lines 7-11). 
Then, we evaluate each adjacent community \( \mathcal{C}_k \) to determine if moving node \( x \) to any of these communities would result in a greater heuristic function value (lines 12-17). 
For each \( \mathcal{C}_k \), we compute the transfer heuristic \( \Delta_{\text{T}}(x, \mathcal{C}_x, \mathcal{C}_k) \) and compare it to the maximum heuristic function value. 
If moving to an adjacent community \( \mathcal{C}_k \) offers a larger heuristic function value, we update \( \Delta_{\max} \) and set \( t \) to the index of community \( \mathcal{C}_k \) (lines 15-17). 
Finally, we find the optimal index  \( t \) and the maximum heuristic function value \( \Delta_{\max} \).
Suppose the strategy indicates that node \( x \) should move to another community.
We transfer \( x \) from its current community \( \mathcal{C}_x \) to the target community \(\mathcal{C}_t\)  (line 19) and update the statistics of the source and target communities, such as adjusting community volumes and cut edge summations via Equation (\ref{EqUpdateTarget}) (line 20).

At the end of each iteration, if all nodes choose to stay in their current community (i.e., $M=0$), the algorithm is considered converged. 
In practice, we introduce a stopping criterion: if the average of $\Delta_{\text{T}}$ decreases significantly compared to the initial average node entropy, it suggests that further adjustments will have little impact on improving the community structure.
Formally, the stopping criterion is defined as:
\begin{equation} 
 \label{eq:termination}
	 \frac{\Delta_{sum}}{M} \leq \frac{\tau_n}{|V|} \sum_{x \in V} - \frac{d_x}{v_\lambda}  \cdot \log \frac{d_x}{v_\lambda} ,
    \end{equation}
where $\Delta_{sum}$ represents the change in the graph's 2D SEduring the current iteration, $M$ denotes the number of nodes that changed communities, and $\tau_n$ is a hyper-parameter, with a range of $(0, 1)$. 
Ultimately, our algorithm outputs the partition \( \mathcal{P} \), which represents the final assignment of nodes into non-overlapping communities. 

\textbf{Overlapping Community Detection.}\label{SecOverlappingDetection} 
The overlapping community detection algorithm begins with a non-overlapping partition \(\mathcal{P}\) of the network \(G\). 
The goal of the overlapping community detection (Algorithm \ref{AlgOverlapping}) is to generate a set of overlapping communities \(\mathcal{P}^o\) of \(G\). 
Initially, \(\mathcal{P}^o\) is identical to \(\mathcal{P}\). For each node \(x\) in the network, the algorithm iterates over all its adjacent communities \(\mathcal{C}_k\). 
It computes the overlapping heuristic function \( \Delta_{\text{O}}(x, \mathcal{C}_k) \). If the heuristic function exceeds the overlap threshold \(\tau_o\), we overlap node \(x\) to community \(\mathcal{C}_k\). 
We define the overlap threshold \(\tau_o\) as the average of node heuristic function values:
\begin{equation} 
	\tau_o = \frac{\gamma}{|\mathcal{C}_k|} \sum_{x_i \in \mathcal{C}_k } { - \Delta_L(x_i, \mathcal{C}_k) },
\end{equation}
where $\gamma$ represents the overlap factor, which is employed to actively regulate the degree of community overlapping. 
The iterative process ensures that nodes overlap in communities, significantly reducing the network's 2D structural entropy. 
\begin{algorithm}[tp]
    \SetAlgoShortEnd
    \caption{Overlapping community detection.}
    \label{AlgOverlapping}
    \let\oldnl\nl
    \newcommand{\nonl}{\renewcommand{\nl}{\let\nl\oldnl}}

    \DontPrintSemicolon
    \KwIn{Non-overlapping communities $\mathcal{P}$.}
    \KwOut{Overlapping communities $\mathcal{P}^o$ }
    
    $\mathcal{P}^o \gets \mathcal{P}$ \;
    
    \For{node $x$ in $\mathcal{V}$}{
        \For{$k$-th adjacent community $\mathcal{C}_k$ of node $x$}{
            $\Delta_{\text{O}}(x, \mathcal{C}_k) \gets $ Eq.\ref{EqOverlap} \;
                    
            \If{$\Delta_{\text{O}}(x, \mathcal{C}_k) > \tau_o $}{
                Overlap node $x$ to $\mathcal{C}_k$ in $\mathcal{P}^o$ \;
            }
        }
    }
    \Return{$\mathcal{P}^o$}
\end{algorithm}

\subsection{\textcolor{black}{Dynamic Community Detection Algorithm}}
\label{Secdynamic}
\begin{figure*}[th]
    \centering
    \includegraphics[width=1\linewidth,trim={0.2cm 0cm 1.4cm 0.1cm}, clip]{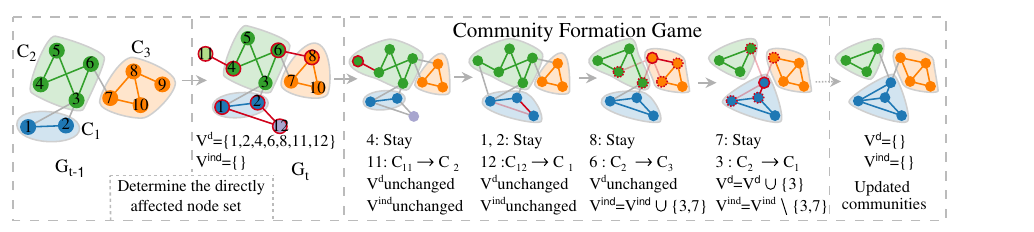}
    \vspace{-4mm}
    \caption{\textcolor{black}{Overview of the CIPA.}\vspace{-4mm}}
    \label{CIPA}
\end{figure*}

\textcolor{black}{As illustrated in Fig.\ref{CIPA}, we present a Cascading Influence Propagation-based Adaptive (CIPA) strategy to extend \framework~for dynamic community detection.
CIPA is derived from the node structural entropy game and adaptively maintains the set of active (gameable) nodes by tracking variations in node-selection strategies. In this way, it preserves detection accuracy while improving computational efficiency.
The proposed method integrates two key components: Hierarchical affected node identification and Adaptive node removal.}

\textbf{\textcolor{black}{The hierarchical affected node identification mechanism}} \textcolor{black}{is designed to dynamically identify nodes whose strategies may change in the current network snapshot. 
To achieve this, the affected nodes are partitioned into two disjoint sets: 
(i) the directly affected set $\mathcal{V}^{d}$, consisting of newly added nodes and those involved in edge additions or deletions, 
and (ii) the indirectly affected set $\mathcal{V}^{ind}$, comprising nodes whose neighbors belong to the directly affected set and whose community memberships have changed. 
Initially, the directly affected set is constructed by quantifying the variations in nodes and edges between two consecutive snapshots:}
\begin{equation}\label{dr_nodes}
    \begin{aligned}
\mathcal{V}^{\text{add}} &= \mathcal{V}_t \setminus \mathcal{V}_{t-1}, \quad \mathcal{E}^{\text{add}} = \mathcal{E}_t \setminus \mathcal{E}_{t-1}, \\
\mathcal{V}^{\text{del}} &= \mathcal{V}_{t-1} \setminus \mathcal{V}_t, \quad \mathcal{E}^{\text{del}} = \mathcal{E}_{t-1} ,\setminus \mathcal{E}_t,\\
\mathcal{V}^{d} &= \mathcal{V}^{\text{add}} \setminus \mathcal{V}^{\text{del}}  \cup  \bigcup_{x \in \mathcal{V}^{\text{del}}} \mathcal{N}_x\\
&\cup \{ x \mid \exists e \in \mathcal{E}^{\text{del}} \cup \mathcal{E}^{\text{add}} , x \in e \},
\end{aligned}
\end{equation}  
\textcolor{black}{where $\mathcal{V}_t$ and $\mathcal{E}_t$ denote the node and edge sets of the current snapshot $G_t$, and $\mathcal{V}_{t-1}$ and $\mathcal{E}_{t-1}$ denote those of the previous snapshot $G_{t-1}$.
The sets $\mathcal{V}^{\text{add}}$ and $\mathcal{E}^{\text{add}}$ represent the newly added nodes and edges, while $\mathcal{V}^{\text{del}}$ and $\mathcal{E}^{\text{del}}$ denote the removed nodes and edges, respectively. 
$\mathcal{N}_x$ denotes a node's neighboring set.
Newly added nodes are initialized as independent communities, leading to the following initial partition: }
\begin{equation}\label{init_patition}
    \mathcal{P}_t = \mathcal{P}_{t-1} \cup \left\{ \{x\} \mid x \in \mathcal{V}^{\text{add}} \right\},
\end{equation}
\textcolor{black}{where $\mathcal{P}_t$ and $\mathcal{P}_{t-1}$ denote the community partitions of $G_{t}$ and $G_{t-1}$, respectively. 
Furthermore, during the game process, if an affected node $x$ ($x \notin \mathcal{V}^{\text{add}}$) updates its strategy (i.e., $S(x) \neq \Delta_s$), all of its previously unaffected neighbors are incorporated into the indirectly affected node set:}
\begin{equation}\label{eqVind}
    \mathcal{V}^{ind} \gets \mathcal{V}^{ind} \cup \{y \mid y \in \mathcal{N}_x \setminus \mathcal{V}^d\}.
\end{equation}

\textbf{\textcolor{black}{The adaptive node removal mechanism}} \textcolor{black}{aims to exclude stabilized nodes from the affected set, thereby reducing redundant computations and enhancing detection efficiency.
Considering the varying degrees of influence, distinct removal strategies are applied to directly and indirectly affected nodes.
Specifically, a directly affected node is deemed stabilized and removed from the affected set if its chosen strategy remains unchanged for $r$ consecutive iterations (i.e., $S(x) = \Delta_s$).
For indirectly affected nodes, those that remain in their current communities are considered unaffected and are consequently removed. 
Conversely, if an indirectly affected node changes its strategy, it is regarded as genuinely influenced and is transferred to the directly affected set for continued monitoring.
This process is formalized as follows:}
\begin{equation}
\begin{cases}
\mathcal{V}^{d} \leftarrow \mathcal{V}^{d} \setminus \{x\} & \text{if } x \in \mathcal{V}^{d} \land l_x \geq r \\
\mathcal{V}^{\text{ind}} \leftarrow \mathcal{V}^{\text{ind}} \setminus \{x\} & \text{if } x \in \mathcal{V}^{\text{ind}} \land S(x) = \Delta_S \\
\begin{cases}
\mathcal{V}^{\text{ind}} \leftarrow \mathcal{V}^{\text{ind}} \setminus \{x\} \\
\mathcal{V}^{d} \leftarrow \mathcal{V}^{d} \cup \{x\}
\end{cases} & \text{if } x \in \mathcal{V}^{\text{ind}} \land S(x) \neq \Delta_S
\end{cases} ~,
\end{equation}
\textcolor{black}{where $l_x$ denotes the number of consecutive iterations during which $x$ has remained stable.
It is worth noting that, for overlapping dynamic networks, one may first obtain the non-overlapping community structure via CIPA, and subsequently derive the overlapping communities based on Algorithm~\ref{AlgOverlapping}.
The pseudo-code of CIPA is provided in \textbf{Appendix D}.}

\subsection{\textcolor{black}{Algorithm Complexity}}
\label{SecTimeComplexity}
\textbf{\textcolor{black}{Time complexity.}}
\textcolor{black}{In static community detection, \framework's time complexity~is \(O(I \cdot d_{\max} \cdot |\mathcal{V}|)\), where \(|\mathcal{V}|\) denotes the number of nodes in the graph, \(d_{\max}\) represents the maximum degree of any node, and \(I\) is the number of iterations required for the algorithm to converge to a stable partition.
In dynamic detection, the time complexity is reduced to \( O(I\cdot d_{\max} \cdot |\mathcal{V}^{d} \cup \mathcal{V}^{ind}|) \), where \( \mathcal{V}^{d} \) and \( \mathcal{V}^{ind} \) denote the sets of directly and indirectly affected nodes, respectively.
Detailed time complexity analysis is provided in \textbf{Appendix E}.
}

\textbf{\textcolor{black}{Space complexity.}}
\textcolor{black}{The overall space complexity of the proposed algorithm is $O(|\mathcal{E}|)$, where $|\mathcal{E}|$ represents the number of edges in the network. 
Specifically, the storage of the node adjacency list requires $O(2 \cdot |\mathcal{E}|) = O(|\mathcal{E}|)$ space, while the space complexities for storing the community volume, community cut, community partition, and node degree are all $O(|\mathcal{V}|)$. 
Consequently, the overall space complexity of the algorithm is $O(|\mathcal{E}| + 4 \cdot |\mathcal{V}|) = O(|\mathcal{E}|)$, where $|\mathcal{V}| \leq |\mathcal{E}|$.}

\subsection{Parallelism Implementation}
\label{SecParallel}

Parallelization can fully exploit the multi-core architecture of modern CPUs, significantly reducing the algorithm's runtime. 
The proposed algorithm is readily parallelizable. 
In our parallelized \framework~algorithm, each node in different threads independently calculates the optimal strategy based on the current partition. During the execution of the movement strategy, a mutex lock ensures the correct update of statistics such as \(C_k\) and \(g_c\).
It is crucial to note that when the volume and cut of a community are concurrently updated by multiple threads, the increments must be recalculated. 
However, it is worth mentioning that as the algorithm progresses, the number of nodes requiring movement decreases sharply, thereby enhancing the acceleration effect of parallelism. 

During the overlapping community detection, \framework~is inherently parallelizable.
A node’s inclusion in an overlapping community is determined by the change in entropy resulting from its addition, which depends only on the outcomes of non-overlapping community detection and the node’s connections to those communities. 
This allows the overlapping strategies for each node to be calculated concurrently.

%% file: 5_experimental_setup.tex
This section provides a comprehensive description of the experimental setup, including the datasets, baseline methods for comparison, evaluation metrics, and implementation details. 

\begin{table*}[t]
    \centering
    \caption{ \textcolor{black}{Statistics of networks.}}
    \label{TabSnapDatasets}    
    \begin{tabularx}{\linewidth}{
    >{\raggedleft\arraybackslash}X |
    *{4}{>{\raggedleft\arraybackslash}X} |
    *{5}{>{\centering\arraybackslash}X}
}
    
        \toprule
        Dataset & \#Nodes & \#Edges & Avg. Deg &  \#Cmty & Overlapping& Directed & Weighted & Dynamic & \#Snapshot \\
        \midrule
        Amazon & 334,863 & 925,872 & 5.530 & 75,149 & \ding{51} & \ding{55} & \ding{55} & \ding{55} & - \\ 
        YouTube & 1,134,890 & 2,987,624 & 5.265 & 8,385 & \ding{51} & \ding{55} & \ding{55} & \ding{55} & - \\ 
        DBLP & 317,080 & 1,049,866 & 6.622 & 13,477 & \ding{51} & \ding{55} & \ding{55} & \ding{55} & - \\ 
        LiveJournal & 3,997,962 & 34,681,189 & 17.349 & 287,512 & \ding{51} & \ding{55} & \ding{55} & \ding{55} & - \\ 
        Orkut & 3,072,441 & 117,185,083 & 76.281 & 6,288,363 & \ding{51} & \ding{55} & \ding{55} & \ding{55} & - \\ 
        Friendster & 65,608,366 & 1,806,067,135 & 55.056 & 957,154 & \ding{51} & \ding{55} & \ding{55} & \ding{55} & - \\ 
        Wiki & 1,791,489 & 28,511,807 & 15.915 & 17,364 & \ding{51} & \ding{51} & \ding{55} & \ding{55} & - \\ 
        \midrule
        X12\_static & 68,841 & 10,141,672 & 294.640 & 503  & \ding{55} & \ding{55} & \ding{51} & \ding{55} & - \\ 
        X18\_static & 64,516 & 17,926,784 & 555.731 & 257 & \ding{55} & \ding{55} & \ding{51} & \ding{55} & - \\ 
        LFR\_50K & 50,000 & 1,255,720 & 50.229 & 871 & \ding{55} & \ding{55} & \ding{51} & \ding{55} & - \\ 
        LFR\_100K & 100,000 & 2,515,416 & 50.308 & 1277 & \ding{55} & \ding{55} & \ding{51} & \ding{55} & - \\ 
        LFR\_200K & 200,000 & 4,971,923 & 49.719 & 1752 & \ding{55} & \ding{55} & \ding{51} & \ding{55} & - \\ 
        LFR\_500K & 500,000 & 12,108,408 & 48.434 & 2241 & \ding{55} & \ding{55} & \ding{51} & \ding{55} & - \\ 
        LFR\_1M & 1,000,000 & 25,566,093 & 51.132 & 2697 & \ding{55} & \ding{55} & \ding{51} & \ding{55} & - \\ 
        \midrule
        X12 & 68,841 & 10,141,672 & 294.640 & 503  & \ding{55} & \ding{55} & \ding{51} & \ding{51} & 10 \\ 
        X18 & 64,516 & 17,926,784 & 555.731 & 257 & \ding{55} & \ding{55} & \ding{51} & \ding{51} & 8 \\ 
        \bottomrule
    \end{tabularx}
\end{table*}

\textbf{Datasets.}\label{sec:Datesets}
To thoroughly evaluate the effectiveness of \framework, we conduct systematic experiments on three types of network datasets:
(1) overlapping community detection on seven large-scale real-world networks from SNAP~\cite{snapnets}, including Amazon, YouTube, DBLP, LiveJournal, Orkut, Friendster, and Wiki;
(2) comparative overlapping community detection on two complex social networks and five large-scale synthetic networks, specifically X12\_static (Tweet12\_static)~\cite{mcminn2013building}, X18\_static (Tweet18\_static)~\cite{mazoyer2020french}, LFRs; 
and (3) \textcolor{black}{dynamic community detection performance evaluation on two real-world dynamic networks, namely X12 (Tweet12) and X18 (Tweet18)}.
Statistics of datasets are summarized in Table~\ref{TabSnapDatasets}, with detailed descriptions provided in \textbf{Appendix F}.

\textbf{Baselines.}
\label{sec:baseline}
To assess the performance of \framework~ across various networks, we compare it with four overlapping detection methods--\textbf{SLPA}~\cite{xie2011slpa},\textbf{Bigclam}~\cite{yang2013overlapping}, \textbf{NcGame}~\cite{ferdowsi2022detecting}, and \textbf{Fox}~\cite{lyu2020fox},  four non-overlapping detection methods--\textbf{Louvain}~\cite{blondel2008fast}, \textbf{DER}~\cite{kozdoba2015community}, \textbf{Leiden}~\cite{traag2019louvain}, and \textbf{FLPA}~\cite{traag2023large}, and \textcolor{black}{four dynamic detection methods--\textbf{QCA}~\cite{nguyen2011adaptive}, \textbf{DynaMo}~\cite{zhuang2019dynamo}, \textbf{DCDME}~\cite{sun2022dynamic}, and \textbf{DCDBFE}~\cite{anuar2025bird}.} 
Detailed descriptions and implementations of baselines are provided in \textbf{Appendix G}.

\textbf{Evaluation Metrics.}\label{Sec:Metrics}
To validate the effectiveness of the proposed algorithm, we employ two widely used evaluation metrics:
the Average F1-Score (F1) \cite{yang2013overlapping,lutov2019accuracy} and Overlapping Normalized Mutual Information (ONMI) \cite{Lancichinetti2009Detecting}. 
For non-overlapping community detection tasks, we instead adopt the standard Normalized Mutual Information (NMI). 
More details on these metrics are provided in \textbf{Appendix H}.

\begin{table*}[t]
    \setlength{\tabcolsep}{5.7pt}
   \centering
    \begin{threeparttable}
        \caption{Results on overlapping networks (\%). The best results are bolded, and the second-best results are underlined. * indicates the results when treating directed networks as undirected. N/A indicates the runtime extended beyond a week.}
        \label{tab:Result1}
        \begin{tabularx}{\linewidth}{l|*{13}{>{\centering\arraybackslash}c|}c}
        \toprule
        Dataset&\multicolumn{2}{c|}{Amazon}&\multicolumn{2}{c|}{YouTube}&\multicolumn{2}{c|}{DBLP}&\multicolumn{2}{c|}{LiveJournal}&\multicolumn{2}{c|}{Orkut}&\multicolumn{2}{c|}{Friendster}&\multicolumn{2}{c}{Wiki}\\
        \midrule
         Metric&ONMI&F1&ONMI&F1&ONMI&F1&ONMI&F1&ONMI&F1&ONMI&F1&ONMI&F1\\
        \midrule
        SLPA    & $9.05$ & $34.37$ & $4.58$ & $26.16$ & $4.45$ & $23.27$ & $2.55$ & $22.29$ & $0.12$ & $8.51$ & $0.03$ & $2.34$ & $\underline{0.48}^*$ & $\underline{12.16}^*$ \\ 
        Bigclam & $7.62$ & $33.30$ & $1.47$ & $27.78$ & $5.96$ & $24.91$ & $2.05$ & $10.67$ & $0.10$ & $11.72$ & N/A & N/A &  N/A & N/A \\
        NcGame  & $\underline{9.94}$ & $\mathbf{36.01}$ & $1.23$ & $11.37$ & $4.89$ & $\underline{23.94}$ & $1.33$ & $7.96$ & $0.07$ & $2.20$ & $0.19$ & $0.10$ & ${0.02}^*$ & ${10.96}^*$ \\ 
        Fox  & $8.88$ & $29.04$ & $\underline{6.67}$ & $\underline{31.77}$ & $\mathbf{7.34}$ & $23.85$ & $\mathbf{4.18}$ & $\underline{26.79}$ & $\underline{0.47}$ & $\underline{24.07}$ & $\underline{0.56}$ & $\underline{19.03}$ & ${0.45}^*$ & ${10.80}^*$ \\ 
        
        \framework & $\mathbf{10.75}$ & $\underline{34.51}$ & $\mathbf{8.17}$ & $\mathbf{36.80}$ & $\underline{7.21}$ & $\mathbf{25.34}$ & $\underline{3.75}$ & $\mathbf{28.46}$ & $\mathbf{0.49}$ & $\mathbf{25.26}$ & $\mathbf{0.73}$ & $\mathbf{19.21}$ & $\mathbf{2.28}$ & $\mathbf{18.92}$ \\
        \midrule
        
        Improve & $\uparrow0.81$ & $\downarrow1.50$ & $\uparrow1.50$ & $\uparrow5.03$ & $\downarrow0.13$ & $\uparrow1.23$ & $\downarrow0.43$ & $\uparrow1.67$ & $\uparrow0.02$ & $\uparrow1.19$ & $\uparrow0.17$ & $\uparrow0.18$ & $\uparrow1.83$ & $\uparrow6.76$ \\ 
        \bottomrule
        \end{tabularx}
    \end{threeparttable}
    \vspace{-4mm}
\end{table*}

\textbf{Implementation.}\label{Implementation}
In our \framework, the termination threshold $\tau_n$, overlapping factor $\gamma$, and round stability threshold $r$ are set to 0.3, 1, and 2, respectively.
All experiments are conducted on a server with a hardware configuration of dual 16-core Intel Xeon Silver 4314 processors @ 2.40GHz and 1024GB of memory.

%% file: 6_experiments.tex
In this section, we conduct extensive experiments to validate the effectiveness of the proposed \framework.
Our goal is to address the following five key questions:
\textbf{RQ1}: How does the \framework~perform in overlapping, non-overlapping, and dynamic community detection tasks compared to baselines?  
\textbf{RQ2}: How does the detection efficiency of \framework~on different tasks compare to the baselines?  
\textbf{RQ3}: How does the performance of the CIPA-based dynamic \framework~compare with that of the static \framework?
\textbf{RQ4}: Can \framework~achieve fast convergence, and how do key hyperparameters impact its performance?  
\textbf{RQ5}: How does parallelization of \framework~influence its performance and efficiency?

\begin{table*}[t]
    \setlength{\tabcolsep}{5.5pt}
   \centering
    \begin{threeparttable}
        \caption{\textcolor{black}{Results on non-overlapping networks (\%). The best results are bolded, and the second-best results are underlined.  OOM indicates Out-Of-Memory.}}
        \label{tab:Result2}  
        \begin{tabularx}{\linewidth}{l|*{13}{>{\centering\arraybackslash}c|}c}
        \toprule
        Dataset&\multicolumn{2}{c|}{X12\_static}&\multicolumn{2}{c|}{X18\_static}&\multicolumn{2}{c|}{LFR\_50K}&\multicolumn{2}{c|}{LFR\_100K}&\multicolumn{2}{c|}{LFR\_200K}&\multicolumn{2}{c|}{LFR\_500K}&\multicolumn{2}{c}{LFR\_1M}\\
        \midrule
         Metric&NMI&F1&NMI&F1&NMI&F1&NMI&F1&NMI&F1&NMI&F1&NMI&F1\\
        \midrule
        Louvain & $52.93$ & $39.63$ & $47.85$ & $53.60$ & $78.34$ & $52.91$ & $80.21$ & $57.63$ & $82.65$ & $63.31$ & $86.65$ & $72.39$ & $87.12$ & $73.69$ \\ 
        DER & $10.62$ & $19.20$ & $11.94$ & $32.76$ & $10.17$ & $5.66$ & $9.55$ & $5.25$ & OOM & OOM & OOM & OOM & OOM &  OOM\\
        Leiden  & $54.37$ & $42.01$ & $48.81$ & $52.13$ & $78.39$ & $52.96$ & $79.70$ & $56.63$ & $82.16$ & $62.26$ & $87.09$ & $73.39$ & $88.68$ & $77.18$ \\
        FLPA  & $\underline{79.39}$ & $\underline{65.14}$ & $\underline{49.25}$ & $\underline{65.05}$ & $\underline{89.77}$ & $\underline{78.45}$ & $\underline{93.03}$ & $\underline{84.38}$ & $\underline{94.46}$ & $\underline{87.48}$ & $\underline{96.11}$ & $\underline{91.03}$ & $\underline{96.08}$ &  $\underline{91.14}$ \\ 
        \framework &  $\mathbf{83.41}$ & $\mathbf{66.61}$ & $\mathbf{77.65}$ & $\mathbf{68.81}$ & $\mathbf{92.99}$ & $\mathbf{89.10}$ & $\mathbf{95.51}$ & $\mathbf{93.61}$ & $\mathbf{96.93}$ & $\mathbf{95.72}$ & $\mathbf{97.86}$ & $\mathbf{97.39}$ & $\mathbf{97.85}$ & $\mathbf{97.70}$ \\ 
        \midrule
        Improve & $\uparrow4.02$ & $\uparrow1.47$ & $\uparrow28.37$ & $\uparrow3.76$ & $\uparrow3.22$ & $\uparrow10.65$ & $\uparrow2.48$ & $\uparrow9.23$ & $\uparrow2.47$ & $\uparrow8.24$ & $\uparrow1.75$ &  $\uparrow6.36$ & $\uparrow1.77$ & $\uparrow6.56$ \\ 
        \bottomrule
        \end{tabularx}
    \end{threeparttable}
\end{table*}

\begin{table*}[t]
    \setlength{\tabcolsep}{5.4pt}
   \centering
    \begin{threeparttable}
        \caption{\textcolor{black}{Results on dynamic networks (\%). The best results are bolded, and the second-best results are underlined.}}
        \label{tab:Result3}  
        \begin{tabularx}{\linewidth}{l|*{13}{>{\centering\arraybackslash}c|}c}
        \toprule
         Dataset& \multicolumn{14}{c}{X12} \\
        \midrule
         Snapshot&\multicolumn{2}{c|}{$G_0$}&\multicolumn{2}{c|}{$G_1$}&\multicolumn{2}{c|}{$G_2$}&\multicolumn{2}{c|}{$G_3$}&\multicolumn{2}{c|}{$G_4$}&\multicolumn{2}{c|}{$G_5$}&\multicolumn{2}{c}{$G_6$}\\
        \midrule
         Metric&NMI&F1&NMI&F1&NMI&F1&NMI&F1&NMI&F1&NMI&F1&NMI&F1\\
        \midrule
        QCA & $80.24$ & $68.76$ & $\underline{82.67}$ & $\underline{69.99}$ & $\textbf{83.65}$  & $\textbf{70.86}$ & $\underline{81.96}$ & $\textbf{70.75}$ & $\underline{78.01}$ & $\underline{62.92}$ & $77.23$ & $\underline{60.37}$ & $75.38$ & $\underline{55.88}$ \\ 
        DynaMo & $\underline{80.54}$ & $68.96$ & $80.31$ & $68.92$ & $74.70$ & $59.47$ & $71.79$ & $57.29$ & $64.33$ & $44.06$ & $65.46$ & $46.05$ & $63.23$ & $42.40$ \\
        DCDME  & $78.75$ & $\underline{69.19}$ & $80.98$ & $68.65$ & $75.33$ & $59.08$ & $75.85$ & $58.22$ & $73.17$ & $58.40$ & $72.32$ & $55.16$ & $71.72$ & $52.64$ \\
        DCDBFE  & $77.12$ & $63.28$ & $79.90$ & $62.80$ & $77.28$ & $59.92$ & $78.55$ & $59.42$ & $77.80$ & $57.01$ & $\underline{78.11}$ & $55.97$ & $\underline{78.48}$ & $55.07$ \\ 
        
        \framework &  $\textbf{84.43}$ & $\textbf{75.17}$ & $\textbf{87.85}$ & $\textbf{77.73}$ & $\underline{82.16}$ & $\underline{68.03}$ & $\textbf{83.79}$ & $\underline{68.98}$ & $\textbf{81.99}$ & $\textbf{65.23}$ & $\textbf{82.42}$ & $\textbf{65.12}$ & $\textbf{83.16}$ & $\textbf{65.53}$ \\ 
        \midrule
        Improve & $\uparrow$$3.89$ & $\uparrow$$5.98$ & $\uparrow$$5.18$ & $\uparrow$$7.74$ & $\downarrow$$1.49$ & $\downarrow$$2.83$ & $\uparrow$$1.83$ & $\downarrow$$1.77$ & $\uparrow$$3.98$ & $\uparrow$$2.31$ & $\uparrow$$4.31$ & $\uparrow$$4.75$ & $\uparrow$$4.68$ & $\uparrow$$9.65$ \\ 
        \midrule
        
        Dataset& \multicolumn{8}{c|}{X12} & \multicolumn{6}{c}{X18} \\
        \midrule
         Snapshot&\multicolumn{2}{c|}{$G_7$}&\multicolumn{2}{c|}{$G_8$}&\multicolumn{2}{c|}{$G_9$}&\multicolumn{2}{c|}{Avg. Metric}&\multicolumn{2}{c|}{$G_0$}&\multicolumn{2}{c|}{$G_1$}&\multicolumn{2}{c}{$G_2$}\\
        \midrule
         Metric&NMI&F1&NMI&F1&NMI&F1&NMI&F1&NMI&F1&NMI&F1&NMI&F1\\
        \midrule
        QCA    & $74.27$ & $\underline{53.86}$ & $72.77$ & $51.33$ & $72.17$ & $50.34$ & $77.84$ & $\underline{61.51}$ & $\underline{70.55}$ & $70.46$ & $67.69$ & $69.45$ & $\underline{71.13}$ & $\textbf{71.39}$ \\ 
        DynaMo & $63.32$ & $42.70$ & $61.42$ & $40.12$ & $61.07$ & $39.67$ & $68.62$ & $50.96$ & $70.52$ & $\underline{70.64}$ & $\textbf{72.76}$ & $\textbf{73.75}$ & $69.32$ & $69.19$ \\
        DCDME  & $71.76$ & $51.44$ & $71.83$ & $50.88$ & $71.58$ & $50.50$ & $74.33$ & $57.42$ & $42.22$ & $50.83$ & $55.27$ & $52.63$ & $40.45$ & $55.48$ \\
        DCDBFE & $\underline{78.21}$ & $53.38$ & $\underline{77.88}$ & $\underline{52.21}$ & $\underline{77.49}$ & $\underline{51.49}$ & $\underline{78.08}$ & $57.06$ & $57.55$ & $50.61$ & $64.57$ & $51.27$ & $62.07$ & $58.75$ \\ 
        
        \framework & $\textbf{83.46}$ & $\textbf{65.11}$ & $\textbf{82.97}$ & $\textbf{64.16}$ & $\textbf{82.82}$ & $\textbf{65.86}$ & $\textbf{83.51}$ & $\textbf{68.09}$ & $\textbf{82.45}$ & $\textbf{78.37}$ & $\textbf{80.15}$ & $\underline{71.69}$ &  $\textbf{77.73}$ & $\underline{69.57}$ \\ 
        \midrule
        Improve & $\uparrow$$5.25$ & $\uparrow$$11.25$ & $\uparrow$$5.09$ & $\uparrow$$11.95$ & $\uparrow$$5.33$ & $\uparrow$$14.37$ & $\uparrow$$5.43$ & $\uparrow$$6.58$ & $\uparrow$$11.90$ & $\uparrow$$7.73$ & $\uparrow$$7.39$ & $\downarrow$$2.06$ & $\uparrow$$6.60$ & $\downarrow$$1.82$ \\ 
        \midrule
        
        Dataset& \multicolumn{12}{c}{X18} \\
        \cmidrule{1-13}
         Snapshot&\multicolumn{2}{c|}{$G_3$}&\multicolumn{2}{c|}{$G_4$}&\multicolumn{2}{c|}{$G_5$}&\multicolumn{2}{c|}{$G_6$}&\multicolumn{2}{c|}{$G_7$}&\multicolumn{2}{c}{Avg. Metric}& \multicolumn{2}{c}{}  \\
        \cmidrule{1-13}
         Metric&NMI&F1&NMI&F1&NMI&F1&NMI&F1&NMI&F1&NMI&\multicolumn{1}{c}{F1}&\multicolumn{2}{c}{}\\
        \cmidrule{1-13}
        QCA & $\underline{65.63}$ & $62.89$ & $\underline{63.75}$ & $62.18$ & $59.14$ & $56.31$ & $56.87$ & $51.84$ & $54.83$ & $\underline{50.13}$ & $\underline{63.70}$ & \multicolumn{1}{c}{$61.83$} & \multicolumn{2}{c}{}   \\ 
        DynaMo & $60.80$ & $\underline{64.07}$ & $62.04$ & $\underline{63.83}$ & $54.64$ & $\underline{56.57}$ & $55.86$ & $\underline{54.27}$ & $52.29$ & $49.70$ & $62.28$ & \multicolumn{1}{c}{\underline{$62.88$}} & \multicolumn{2}{c}{}  \\
        DCDME  & $35.86$ & $50.71$ & $44.48$ & $49.94$ & $42.61$ & $46.85$ & $43.09$ & $45.04$ & $42.97$ & $44.12$ & $43.37$ & \multicolumn{1}{c}{$49.45$} & \multicolumn{2}{c}{}   \\
        DCDBFE  & $61.11$ & $53.85$ & $62.67$ & $50.97$ & $\underline{59.75}$ & $46.86$ & $\underline{61.65}$ & $47.89$ & $\underline{62.07}$ & $47.90$ & $61.43$ & \multicolumn{1}{c}{$51.01$} & \multicolumn{2}{c}{}  \\ 
        
        \framework &  $\textbf{75.92}$ & $\textbf{65.75}$ & $\textbf{76.86}$ & $\textbf{66.66}$ & $\textbf{76.33}$ & $\textbf{66.31}$ & $\textbf{77.05}$ & $\textbf{66.98}$ & $\textbf{77.42}$  & $\textbf{67.62}$ & $\textbf{77.99}$ & \multicolumn{1}{c}{$\textbf{69.12}$} &  \multicolumn{2}{c}{}  \\ 
        \cmidrule{1-13}
        Improve & $\uparrow$$10.29$ & $\uparrow$$1.68$ & $\uparrow$$13.11$ & $\uparrow$$2.83$ & $\uparrow$$16.58$ & $\uparrow$$9.74$ & $\uparrow$$15.40$ & $\uparrow$$12.71$ & $\uparrow$$15.35$ & $\uparrow$$17.49$ & $\uparrow$$14.29$ & \multicolumn{1}{c}{$\uparrow$$6.24$} & \multicolumn{2}{c}{} \\
        \cmidrule{1-13}
        \cmidrule{1-13}
        \end{tabularx}
    \end{threeparttable}
    \vspace{-4mm}
\end{table*}

\subsection{Effectiveness of \framework}

In this section, we perform experiments on overlapping, non-overlapping, and dynamic networks to comprehensively evaluate the effectiveness of \framework. 
As summarized in Tables~\ref{tab:Result1} to \ref{tab:Result3}, the results indicate that \framework~significantly outperforms all baseline methods across all evaluated network scenarios, confirming the effectiveness of the utility function that combines potential games with structural entropy in uncovering network community structures.

\subsubsection{Overlapping community detection}
As shown in Tables~\ref{tab:Result1}, \framework~consistently ranks first or second in both ONMI and F1 scores across all seven overlapping networks, demonstrating particularly strong performance on YouTube, Orkut, Friendster, and Wiki, where it achieves the highest values.
This highlights the effectiveness and robustness of using structural entropy to uncover community structures in networks. 
Moreover, the proposed overlapping heuristic function effectively distinguishes the overlapping community memberships of nodes.
Specifically, compared to SLPA and Bigclam, \framework~consistently achieved the highest ONMI and F1 scores across all datasets.
On larger networks such as Orkut and Friendster, the performance of SLPA and Bigclam deteriorated significantly. 
For SLPA, the inherent randomness in the selection of initial nodes and the label propagation process may negatively affect the detection results.
Similarly, Bigclam relies on an initial estimate of community size, and inaccuracies in estimating the number of communities can adversely affect the network's structural analysis.
Although the NcGame algorithm achieves the best F1 score on the Amazon dataset, its generalization ability on other datasets is relatively weak, with performance dropping significantly. 
This limitation arises because NcGame focuses exclusively on maximizing each node’s individual utility, neglecting the constraints that nodes must satisfy when joining communities.
Additionally, the second-best FOX algorithm determines community memberships based on an approximate evaluation of the number of local triangles associated with each node to assess community cohesiveness. This approximation can result in incorrect community assignments, particularly for high-degree nodes.
Notably, \framework~uniquely supports community detection in the large-scale directed network Wiki, whereas baseline methods exhibit significant performance degradation when Wiki is treated as an undirected network. 
Specifically, \framework~improves ONMI by 381.25\% and F1 by 55.59\% compared to the best-performing baseline, SLPA, underscoring its distinct advantages in directed network scenarios.

\subsubsection{Non-overlapping community detection}
\label{NOCD}

As shown in Table~\ref{tab:Result2}, compared to all baselines, \framework~achieves an average improvement of 31.33\% in NMI and 4.02\% in F1 score on X12\_static and X18\_static. 
On large-scale synthetic networks, the average improvements in NMI and F1 score reach 2.51\% and 9.63\%, respectively.
Specifically,  Leiden and Louvain consistently underperform compared to \framework~across all networks. 
This is due to their tendency to excessively merge small communities during modularity maximization—a limitation exacerbated in dense networks. 
DER demonstrates poor performance in large-scale dense networks, primarily because its low-dimensional vectors fail to discriminate community structures under dense connectivity. 
Moreover, the algorithm incurs prohibitively high space complexity, exceeding available system memory when processing networks with 200K nodes (LFR\_200K).
Although FLPA works well on synthetic networks with clear community structures, its effectiveness declines on dense real-world networks as accumulated randomness during label propagation destabilizes community partitioning.
In contrast, \framework~exhibits significantly higher stability and robustness.
Visual analysis of detection results can be found in \textbf{Appendix I}.

\subsubsection{\textcolor{black}{Dynamic community detection}}
\label{DCDE}
\textcolor{black}{
As shown in Table~\ref{tab:Result3}, \framework~achieves significant improvements: NMI and F1-scores increase by 6.95\% and 10.70\% on X12, and by 22.43\% and 9.92\% on X18, respectively. 
This demonstrates the strong adaptability of the CIPA-based \framework~to complex dynamic networks.
For the DCDME and DCDBFE methods, their reliance solely on local topological similarity while ignoring edge weights leads to significant accuracy degradation in dense networks with ambiguous community boundaries.
For modularity-based methods (QCA and DynaMo), they suffer from the resolution limit of modularity, which biases results toward merging small communities.
Specifically, QCA relies on predefined rules for the local modularity optimization of atomic events (single-node/edge changes). 
Over the long-term evolution of the network, the bias from these local decisions accumulates and eventually induces drift in community structure. 
Although DynaMo further refines the classification of network evolution events, its incremental update strategy only covers directly affected nodes and their first-order neighbors, failing to detect all potentially affected nodes transitively.
Notably, the CIPA strategy employs a hierarchical affected node identification mechanism that more effectively covers globally affected nodes. 
Moreover, the utility function derived from community potential game and the network's 2D SE enables nodes to make globally optimal decisions, resulting in more robust outcomes in dynamic community detection.}

\begin{table}[b]
    \vspace{-3mm}
    \centering
    \caption{Time consumption of different algorithms in static networks (Sec). OOM indicates Out Of Memory.}
    \label{tab: Time consumption}  
    \setlength{\tabcolsep}{3pt}
    \begin{tabularx}{\linewidth}{l|ccccc|r}
        \toprule
        \multicolumn{7}{c}{Overlapping Networks}\\
        \midrule
        Method & Bigclam & SLPA & NcGame & Fox & \framework & Ratio \\
        \midrule
        Amazon      & 482.2 & 368.7 & 44.0 & \underline{15.2} & \textbf{2.6} & 5.8 \\
        YouTube     & 442764.0 & 852.4 & 1374.1 & \underline{219.7} & \textbf{5.9} & 36.9 \\
        DBLP        & 2106.2& 263.4 & 62.1 & \underline{15.8} & \textbf{2.3} & 6.9\\
        LiveJournal & \underline{1064.3} & 11343.1 & 4146.8 & 2330.2 & \textbf{51.4} & 20.7 \\
        Orkut       & \underline{2898.7} &21429.0 & 37590.2 & 73385.5& \textbf{278.8} & 10.4 \\
        Friendster  & $>7$day& 298536.2& \underline{191556.1}& 487396.9& \textbf{5650.3} & 33.9 \\
        \midrule
        Wiki        & $>7$day& \underline{4410.0} & 36233.1 & 15511.5 & \textbf{27.8} & 158.6 \\
        \midrule
        \multicolumn{7}{c}{Non-overlapping Networks}\\
        \midrule
        Method & Louvain & DER & Leiden & FLPA & \framework & Ratio \\
        \midrule
        X12\_static & 26.8 & 3675.7 & 65.7 &\underline{21.4}  &\textbf{12.7} &1.7\\
        X18\_static & 48.3 & 7622.1 & 151.2 &\underline{50.3} &\textbf{15.3} &3.3 \\
        LFR\_50K & 2.3 & 905.0 & 12.9 &\textbf{0.5} &\underline{0.6} & 0.8 \\
        LFR\_100K & 4.7 & 4188.4 & 25.4 &\textbf{1.1} &\underline{1.4} &0.8 \\
        LFR\_200K & 9.5 & OOM & 47.5 &\underline{4.3} &\textbf{3.9} &1.1 \\
        LFR\_500K & 26.7 & OOM & 114.7 &\underline{13.1} &\textbf{10.3} &1.3 \\
        LFR\_1M & 67.4 & OOM & 246.5 &\underline{25.8} &\textbf{22.1} &1.2 \\
        \bottomrule
    \end{tabularx}
\end{table} 
\subsection{Efficiency of \framework}
\label{Effic}
Table~\ref{tab: Time consumption} reports the time consumption on overlapping and non-overlapping networks. 
For overlapping detection, \framework~is consistently the most efficient, with the advantage becoming more pronounced on larger networks (e.g., LiveJournal, Orkut, and Friendster). 
On average, \framework~achieves a $33\times$ speedup over the fastest baseline per network. 
This is mainly because \framework~converges in only a few iterations, whereas Bigclam typically requires many more. 
Moreover, Bigclam exhibits unstable convergence, leading to an unusually long runtime on YouTube (even longer than on larger networks such as LiveJournal and Orkut). 
SLPA's detection time is limited by the specified number of iterations and the need to iteratively compare and remove detected communities, where some communities are subsets of others. 
NcGame cannot dynamically update key variables during detection, introducing additional computation on large-scale graphs. 
For FOX, runtime grows with the number of overlapping nodes.
For non-overlapping detection, \framework~yields a $546\times$ speedup over DER and is on average $1.8\times$ faster than the fastest baseline, FLPA. 
While FLPA is comparable to \framework~on small graphs (LFR\_50K and LFR\_100K), its efficiency degrades progressively as the network scales. 
Overall, \framework's superior performance and efficiency make it highly scalable for large, complex, real-world networks.

\begin{figure}[t]
    \centering
    \includegraphics[width=1\linewidth, trim={0.4cm 3cm 8cm 0.5cm}, clip]{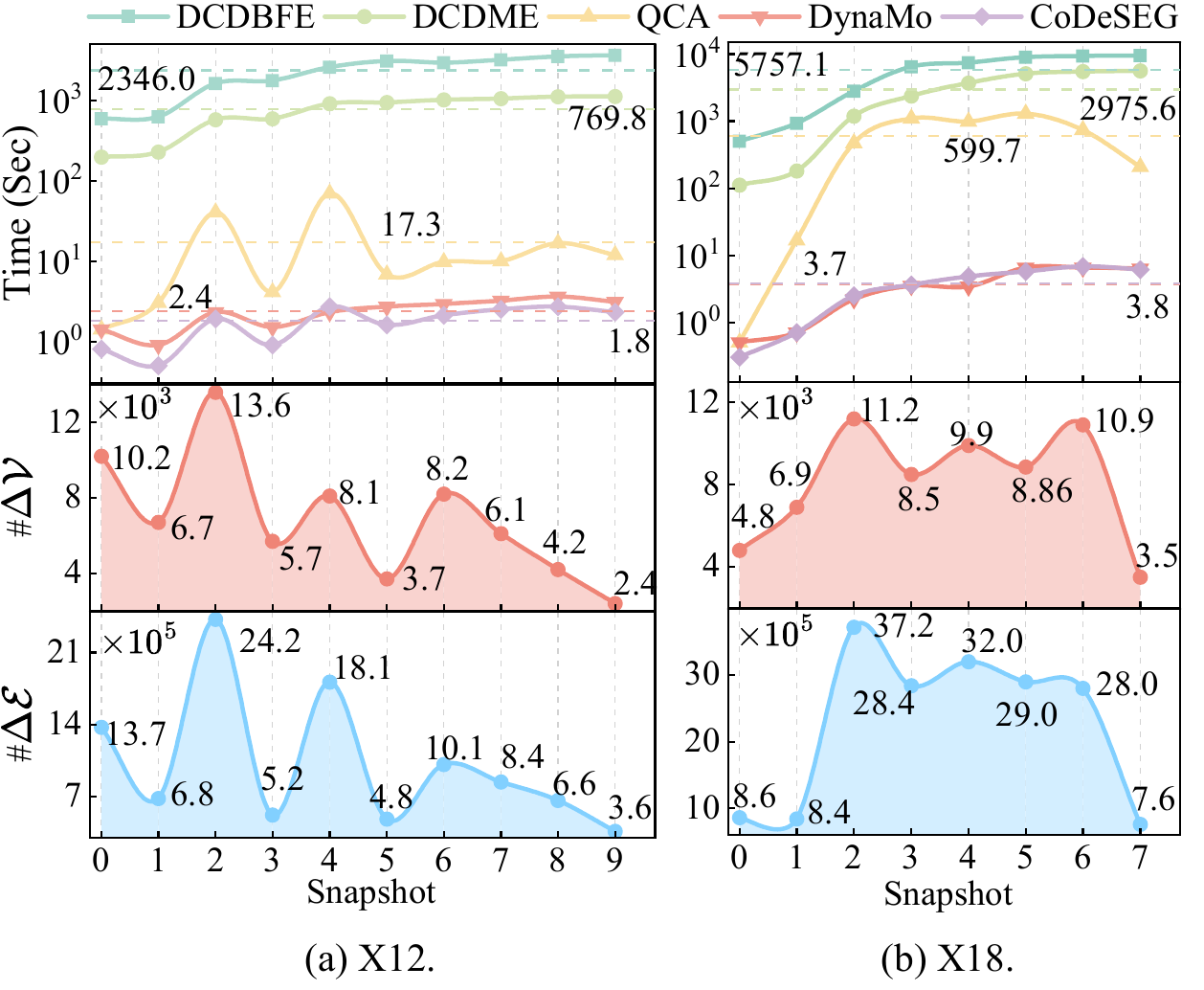}
    \caption{\textcolor{black}{Time consumption of algorithms in dynamic networks.}}
    \label{tab:time_dy}
    \vspace{-3mm}
\end{figure}

As \textcolor{black}{shown in Figure~\ref{tab:time_dy}, we measure snapshot-to-snapshot node/edge changes on two dynamic networks (X12 and X18) and report the runtime of all dynamic methods per snapshot.
\framework~remains competitive efficiency in dynamic detection.
Although DynaMo (the fastest baseline) attains a similar runtime to \framework~on X12 and X18, its detection performance is notably worse. 
Compared with QCA, which achieves the best accuracy, \framework~delivers an $80\times$ speedup, largely due to the CIPA-based adaptive node removal that avoids redundant computations.
DCDME and DCDBFE suffer severe bottlenecks on large-scale dynamics: their utility function computations are expensive, and they must re-process all potentially affected communities at each snapshot, causing their efficiency to degrade to the level of static detection when applied to rapidly evolving, complex networks.
Overall, \framework~achieves the best trade-off between efficiency and accuracy among all dynamic baselines.}

\subsection{\textcolor{black}{Stability of~\framework~in Dynamic Network}}
\label{SDN}
\begin{figure}[t]
    \centering
    \includegraphics[width=1\linewidth, trim={0.1cm 1.5cm 7cm 0.1cm}, clip]{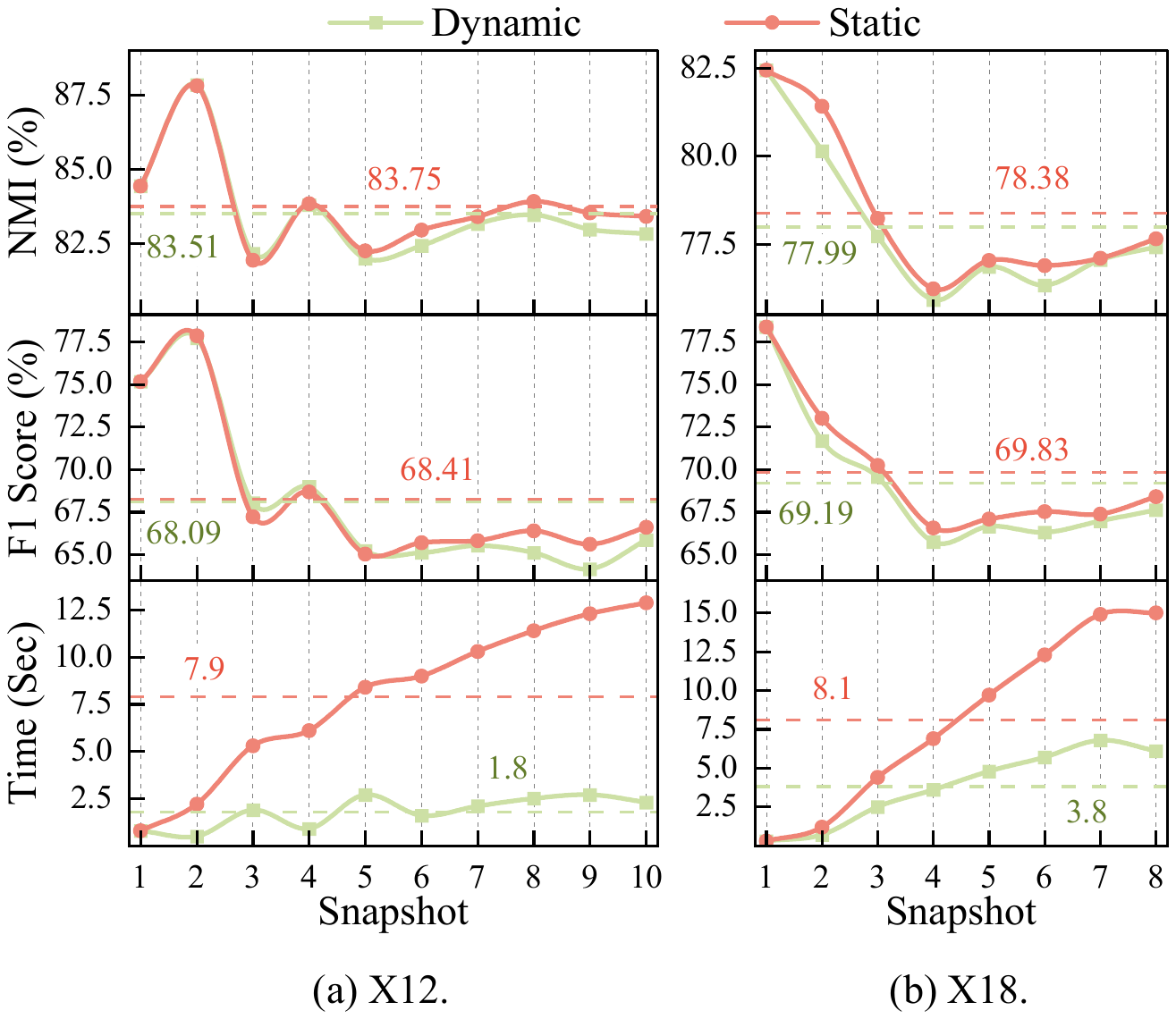}
    \caption{\textcolor{black}{Results on dynamic and static \framework.}\vspace{-3mm}}
    \label{fig:dy_sta}
\end{figure}

\textcolor{black}{To assess stability in dynamic scenarios, we compare the static and dynamic versions of \framework~on X12 and X18. 
Figure~\ref{fig:dy_sta} reports the results, where `Dynamic' denotes the CIPA-extended \framework~and `Static' reruns the original \framework~from scratch on each snapshot. 
Across both networks, the dynamic \framework~achieves performance comparable to the static one: on X12, the average NMI/F1 drops are only 0.0024/0.0032, and on X18 they are merely 0.0039/0.0064. 
This consistency stems from CIPA accurately identifying potentially affected nodes, ensuring stable detection results between the dynamic and static \framework. 
Meanwhile, the dynamic \framework~is substantially more efficient, yielding $4.39\times$ and $2.13\times$ speedups on X12 and X18, respectively, by updating only communities of potentially affected nodes per snapshot. 
Overall, the CIPA-extended \framework~adapts well to rapidly evolving networks while maintaining detection quality on par with the static algorithm.}

\subsection{Convergence of~\framework}
\label{Sec:Convergence}

\begin{figure}[t]
    \centering
    \includegraphics[width=1\linewidth, trim={2cm 1.2cm 3cm 0.8cm}, clip]{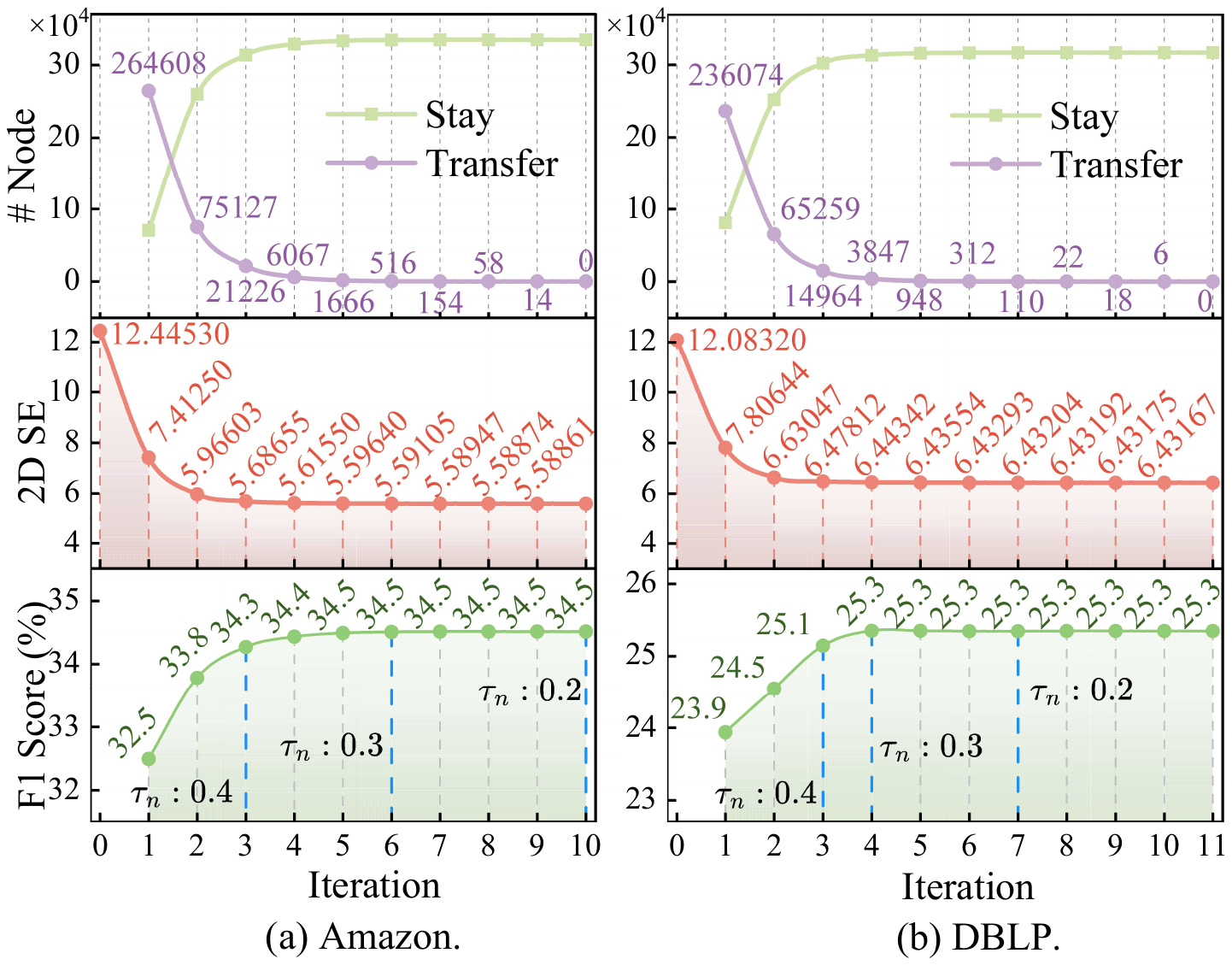}
    \caption{Convergence of ~\framework~ on Amazon and DBLP.\vspace{-3mm}}
    \label{fig:Convergence}
\end{figure}

Figure \ref{fig:Convergence} illustrates the convergence of \framework~during the non-overlapping detection phase on the Amazon and DBLP networks. 
As shown, both the 2D SE of the networks and the number of node movements decrease significantly within the first three iterations. 
By the fifth iteration, the algorithm has essentially converged, with the number of moved nodes accounting for only 1/200 of the total nodes. 
The algorithm's linear complexity and rapid convergence make it highly efficient for large-scale complex networks. 
Furthermore, ~\framework~ demonstrates exceptional robustness and stability, with a smooth convergence process and no repeated node adjustments in the later iterations. 
All nodes successfully join their optimal communities, achieving an accurate Nash equilibrium.
Moreover, as observed from Figure \ref{fig:Convergence}, the F1 score on Amazon stabilizes at the 5th iteration, while the F1 score on DBLP stabilizes at the 4th iteration. 
This indicates that the impact of a small number of unstable nodes on the final partition quality during the non-overlapping detection phase is minimal.
Although a few nodes may not be assigned to their optimal communities in the non-overlapping phase, these nodes can be replicated to other appropriate communities—including their optimal ones—during the subsequent overlapping detection phase.
Furthermore, we investigate the effect of the termination threshold $\tau_n$ on the algorithm's performance. 
If $\tau_n$ is set too high (e.g., 0.4), the algorithm may terminate prematurely, leaving many nodes in an unstable state, thereby compromising performance. Conversely, if $\tau_n$ is too low (e.g., 0.2), the algorithm requires more iterations, increasing computational time. 
By selecting an appropriate  $\tau_n$ (e.g., 0.3), the algorithm can maintain performance while effectively improving detection efficiency.
\begin{figure}[b]
    \vspace{-3mm}
    \centering
    \includegraphics[width=1\linewidth, trim={0.1cm 8cm 8cm 0.5cm}, clip]{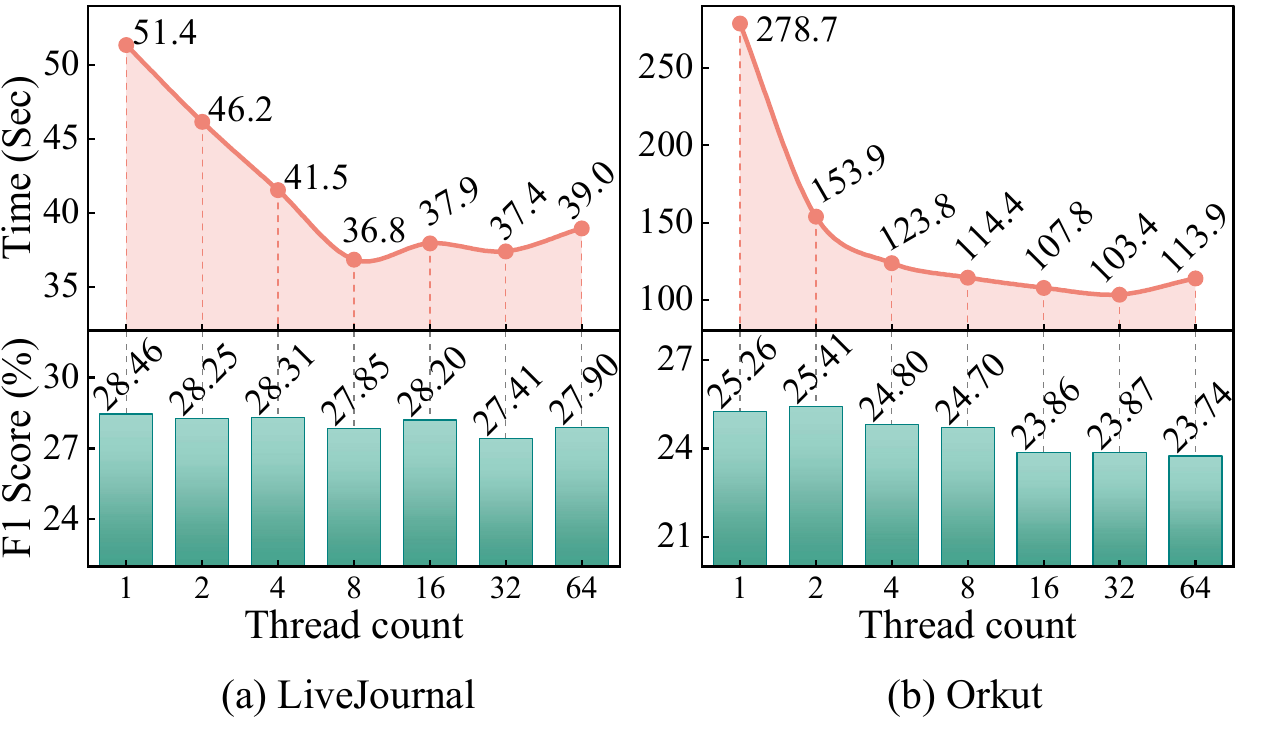}
    \caption{Multithreaded results on LiveJournal and Orkut.}
    \label{fig:thread}
\end{figure}
\subsection{Parallelization Study}
\label{Sec:parallelization}
The primary objective of parallelizing the algorithm is to significantly reduce runtime while preserving as much of the performance as possible. 
To assess the impact of varying thread counts on performance and efficiency, we conduct experiments on the LiveJournal and Orkut networks using different thread configurations, as presented in Figure \ref{fig:thread}. 
The results reveal that although \framework’s partitioning outcomes show minor variations across different thread counts, these differences are negligible compared to the ground truth communities.
Additionally, runtime reduction does not follow a strictly linear pattern as thread count increases. 
For instance, runtime with 32 threads is shorter in both networks than 64 threads.
This can be attributed to the fact that as thread count rises, computational discrepancies between threads may lead to an increase in the number of iterations, thus prolonging the total runtime.
Moreover, with more threads, overhead related to thread initialization, lock contention, and synchronization delays also rise. 
Therefore, balancing task partitioning granularity and the efficiency gains of parallelization is crucial when determining the optimal number of threads.

%% file: 2_related_work.tex
\textcolor{black}{This section provides an overview of the existing literature relevant to the proposed \framework.
The discussion is structured into three categories: Static Community Detection, Dynamic Community Detection, and Structural Entropy.}

\textcolor{black}{\textbf{Static Community Detection.}
Community detection has been studied for over two decades. Representative approaches include modularity optimization~\cite{cherifi2019community,blondel2008fast,traag2019louvain,guo2022local}, label propagation~\cite{lu2018lpanni,xie2011slpa,traag2023large}, seed expansion~\cite{whang2013overlapping,asmi2022greedy,zhao2021community}, non-negative matrix factorization~\cite{yang2013overlapping,luo2020highly,berahmand2022graph}, spectral clustering~\cite{van2019scalable,berahmand2022novel,li2018local}, and deep learning methods~\cite{yang2016modularity, zou2023se, sunlsenet,zeng2024scalable}. 
Deep learning typically performs community clustering/classification via node embeddings, yet the high encoding cost limits its use on large-scale networks~\cite{su2024comprehensive,jin2023Statistical}. 
From a structural viewpoint, methods can be divided into non-overlapping~\cite{blondel2008fast,traag2019louvain,guo2022local, traag2023large,zeng2024scalable} and overlapping community detection~\cite{yang2013overlapping,xie2011slpa,Alvari2011Detecting,lyu2020fox,ferdowsi2022detecting}.
However, many algorithms assume specific graph types or community structures: Louvain~\cite{blondel2008fast} and FLPA~\cite{traag2023large} mainly target disjoint communities in undirected graphs, whereas BigClam~\cite{yang2013overlapping} and SLPA~\cite{xie2011slpa} focus on overlapping communities in undirected graphs.
Recently, game-theoretic algorithms have gained attention for effective and efficient community discovery by modeling interdependent agent decisions. Chen et al. propose a community game that motivates subsequent designs of utility functions for disjoint detection~\cite{chen2010game};
Alvari et al.~\cite{Alvari2011Detecting} leverage structural equivalence to uncover overlapping structures; 
Crampes et al.~\cite{Crampes2015Overlapping} formulate a potential game for node reassignment but require a pre-specified number of communities. 
For large-scale graphs, FOX~\cite{lyu2020fox} estimates node--community closeness by approximately counting intra-community triangles. 
Ferdowsi et al.~\cite{ferdowsi2022detecting} develop a two-stage non-cooperative game that detects disjoint communities via local utilities and then identifies overlapping nodes based on membership benefits. 
Nevertheless, both efficiency and detection quality in large-scale settings still need further improvement. 
In contrast, we propose a utility function based on 2D SE and community potential game that facilitates efficient community detection while accounting for the global partition of the network.} 

 
\textcolor{black}{\textbf{Dynamic Community Detection.}
Many real-world networks evolve continuously, and directly applying static community detection to dynamic networks is often computationally prohibitive, motivating dedicated dynamic community detection methods. Existing approaches are commonly grouped into three classes: \emph{instant-optimal}, \emph{temporal trade-off}, and \emph{incremental} (our method belongs to the incremental class). Instant-optimal methods~\cite{bovet2022flow,ilhan2015predicting,dhouioui2014tracking} rerun static algorithms on affected regions and then match/merge the results with previous communities, but repeated static optimization becomes a major bottleneck in large-scale settings. Temporal trade-off methods~\cite{ma2017evolutionary,rossetti2017tiles} jointly detect and track communities by leveraging both current and historical snapshots, yet they often require pre-specifying the number of communities.
Incremental methods~\cite{seifikar2020c,anuar2025bird,sun2022dynamic,nguyen2011adaptive,zhuang2019dynamo} update community structures by adapting to snapshot-to-snapshot changes. FacetNet~\cite{lin2009analyzing} iteratively discovers dynamic communities by combining non-negative matrix factorization with cost optimization. QCA~\cite{nguyen2011adaptive} uses hand-crafted update rules for node/edge additions and deletions, while DynaMo~\cite{zhuang2019dynamo} enhances detection speed by further segmenting network change events and processing them in batches.DCDME~\cite{sun2022dynamic} and DCDBFE~\cite{anuar2025bird} extract affected subgraphs and revise communities based on the Matthew effect and bird flock effect, respectively. However, as the network evolves, such locality-based updates may miss potentially affected nodes, causing detection quality to degrade over time.
Based on the CIPA strategy, our algorithm achieves a better balance between efficiency and detection quality.}

 
\textcolor{black}{\textbf{Structural Entropy.}
Structural Entropy (SE)~\cite{li2016structural}, based on Shannon's information theory and the Minimum Description Length principle, quantifies the uncertainty of graph structures across hierarchical levels. 
Initially applied to community detection, employing a global greedy strategy to merge communities within a network by minimizing the 2D SE. 
However, as noted in the literature, the original SE minimization process suffers from prohibitively high computational complexity, posing a fundamental challenge to its efficient implementation.
This paper effectively addresses this challenge. 
By integrating community game theory, we successfully reduce the computational complexity of minimizing two-dimensional SE (2D SE) to a linear level.
Furthermore, this study extends the theory of SE to scenarios involving both overlapping community detection and dynamic community detection.
Beyond community detection, structural entropy finds broad application across diverse domains, including social event detection~\cite{cao2024hierarchical,yu2025towards}, social bot detection~\cite{peng2024unsupervised,yang2024sebot}, graph structure learning~\cite{duan2024structural,zou2023se}, and image segmentation~\cite{xie2025hierarchical,,zeng2023unsupervised}.}


%% file: 7_conclusion.tex
In this paper, we propose a fast heuristic community detection algorithm by minimizing the 2D SE of networks within the framework of a potential game. 
By designing a utility function with nearly linear time complexity, our algorithm can efficiently detect high-quality communities in large-scale networks, completing the process within minutes, even for networks comprising millions of nodes. 
Furthermore, we introduce the CIPA strategy, which extends the original algorithm to dynamic community detection. 
This strategy enables the algorithm to efficiently and adaptively update communities in dynamic networks, eliminating the need for full network re-detection. 
Extensive experiments on multiple large-scale networks demonstrate the practical effectiveness and computational efficiency of our approach.
We envision broad applications for \framework~across diverse fields, including social networks, biomedicine, and e-commerce.
In future work, we aim to extend the algorithm to handle more complex heterogeneous networks and further enhance its efficiency through distributed computing paradigms and advanced subgraph partitioning techniques.

%% file: Appendix.tex
\appendix
\subsection{Glossary of Notations}
Notations used in this paper, along with their corresponding description, are presented in Table~\ref{TabNotations}.

\subsection{Structural Entropy}
\label{AppdxStructuralEntropy}
SE~\cite{li2016structural} is defined on an encoding tree, where the encoding tree $\mathcal{T}$ of a graph $G = (\mathcal{V}, \mathcal{E})$ represents a hierarchical partition of $G$ and satisfies the following conditions:
\begin{enumerate} [left=0pt]
    \item Each node $\alpha$ in $\mathcal{T}$ corresponds to a subset of nodes $T_{\alpha} \subseteq \mathcal{V}$. 
    The root node $\lambda$ of $\mathcal{T}$ contains the entire set of nodes, i.e., $T_{\lambda} = \mathcal{V}$. Each leaf node $\gamma$ in $\mathcal{T}$ is associated with exactly one node from the graph $G$, meaning $T_{\gamma} = \{x\}$, where $x \in \mathcal{V}$.

    \item For each node $\alpha$ in $\mathcal{T}$, denote all its children as $\beta_1, \dots, \beta_k$, then $T_{\beta_1}, \dots, T_{\beta_k}$ is a partition of $T_{\alpha}$.

    \item For each node $\alpha$ in $\mathcal{T}$, denote its height as $h(\alpha)$. Let $h(\gamma) = 0$ and $h(\bar{\alpha}) = h(\alpha)+1$, where $\bar{\alpha}$ is the parent of $\alpha$. The height of $\mathcal{T}$ , $h(\mathcal{T}) = \max\limits_{\alpha \in \mathcal{T}}{h(\alpha)}$.
\end{enumerate}

The SE of graph $G$ on encoding tree $\mathcal{T}$ is defined as:
\begin{equation}
    \mathcal{H}^{\mathcal{T}}(G) = - \sum_{\alpha \in\mathcal{T},\alpha \neq \lambda } \frac{g_\alpha}{vol(\lambda)}\log \frac{vol(\alpha)}{vol(\bar{\alpha})},
\end{equation}
where $g_\alpha$ is the summation of the degrees of the cut edges of $\mathcal{T}_\alpha$ (i.e., the weight sum of edges with exactly one endpoint in $\mathcal{T}_\alpha$).
$vol(\alpha)$, $vol(\bar{\alpha})$, and $vol(\lambda)$ represent the volumes, i.e., the sums of node degrees within $\mathcal{T}_\alpha$, $\mathcal{T}_{\bar{\alpha}}$ and $\mathcal{T}_\lambda$, respectively.
The \( d \)-dimensional SE of \( G \) is realized by acquiring an optimal encoding tree of height \( d \), in which the disturbance derived from noise or stochastic variation is minimized:
    \begin{equation}
    \mathcal{H}^{(d)}(G) = \min_{\forall \mathcal{T}: h(\mathcal{T}) = d} \{ \mathcal{H}^{\mathcal{T}} (G) \}.
\end{equation}

Communities within a network can be identified by minimizing its \textbf{2D SE}. Suppose $\mathcal{P}=\{\mathcal{C}_1, \mathcal{C}_2, \dots, \mathcal{C}_k\}$ is a partition of the network $G$, the 2D SE is defined as:
\begin{equation}
    \begin{aligned}
        \mathcal{H}^{2}(\mathcal{P})  
        = - \sum_{c \in \mathcal{P} } \left( \frac{g_{c}}{v_\lambda} \log \frac{v_c}{v_\lambda} + \sum_{x \in c }  \frac{d_x}{v_\lambda} \log \frac{d_x}{v_c} \right),
    \end{aligned}
\end{equation}
where $d_x$ is the degree of node $x$, $v_\lambda$ is the network's volume.
$g_c$ and $v_c$ denote the volume and cut of $c$, respectively.

\subsection{Proof of the $\Delta_{\text{L}}(x, \mathcal{C})$ Computation Formula}
\label{AppdxDeltaLeave}
Suppose the original partition is $\mathcal{P}=\{\mathcal{C}_1, \mathcal{C}_2, \dots, \mathcal{C}_k\}$ and when node $x$ leaves $\mathcal{C}_k$, forming a new partition $\mathcal{P}^\prime=\{\mathcal{C}_1, \mathcal{C}_2, \dots, \mathcal{C}_k^\prime, \{ x\}\}$, where $\mathcal{C}_k = \mathcal{C}_k^\prime \cup \{x\}$.  $\mathcal{T}^\prime$ and $\mathcal{T}$ are encoding tree according to $\mathcal{P}^\prime$ and $\mathcal{P}$, $\lambda$ is the root node of the encoding tree of graph $G$.

\begin{equation}
    \label{EqDeltaLeaveProof}
    \begin{aligned}
    \Delta_L&(x, \mathcal{C}) = \mathcal{H}^{2}(\mathcal{P})-\mathcal{H}^{2}(\mathcal{P}^\prime) \\
        =& - \sum_{\alpha \in \mathcal{T}, \alpha \neq \lambda} \frac{g_\alpha}{v_{\lambda}} \log \frac{v_\alpha}{v_{\bar{\alpha}}} + \sum_{\alpha^\prime \in \mathcal{T}^\prime, \alpha^\prime \neq \lambda} \frac{g_{\alpha^\prime}}{v_\lambda} \log \frac{v_{\alpha^{\prime}}}{v_{\bar{\alpha}^\prime}}  \\
        =& \underbrace{ - \frac{g_{\mathscr{c}_k}}{v_\lambda} \log \frac{v_{\mathscr{c}_k}}{v_\lambda} - \sum_{x_i \in \mathcal{C}_k \setminus \{x\}} \frac{d_{x_i}}{v_\lambda} \log \frac{d_{x_i}}{v_{\mathscr{c}_k}}  }_{\mathcal{H}^{2}(\mathcal{C}_k \setminus \{x\})} 
           \underbrace{ -\frac{d_x}{v_\lambda} \log \frac{d_x}{v_{\mathscr{c}_k}}  }_{\mathcal{H}^{2}(x)}  \\
        & \underbrace{ + \frac{g_{\mathscr{c}_k^\prime}}{v_\lambda} \log \frac{v_{\mathscr{c}_k^\prime}}{v_\lambda} + \sum_{x_i \in \mathcal{C}_k^\prime} \frac{d_{x_i}}{v_\lambda} \log \frac{d_{x_i}}{v_{\mathscr{c}_k^\prime}} }_{-\mathcal{H}^2(\mathcal{C}_k^\prime)} 
               \underbrace{ + \frac{d_x}{v_\lambda} \log \frac{d_x}{v_\lambda}  }_{-\mathcal{H}^2(\{x\})} \\
        =& \frac{g_{\mathscr{c}_k^\prime}}{v_\lambda} \log \frac{v_{\mathscr{c}_k^\prime}}{v_\lambda}  - \frac{g_{\mathscr{c}_k}}{v_\lambda} \log \frac{v_{\mathscr{c}_k}}{v_\lambda} + \frac{d_x}{v_\lambda}(\log \frac{d_x}{v_\lambda} - \log \frac{d_x}{v_{\mathscr{c}_k}} ) \\
        &+ \sum_{x_i \in \mathcal{C}_k^\prime} \frac{d_{x_i}}{v_\lambda} ( \log \frac{d_{x_i}}{v_{\mathscr{c}_k^\prime}} - \log \frac{d_{x_i}}{v_{\mathscr{c}_k}}) \\
        =& \frac{g_{\mathscr{c}_k^\prime}}{v_\lambda} \log \frac{v_{\mathscr{c}_k^\prime}}{v_\lambda}  - \frac{g_{\mathscr{c}_k}}{v_\lambda} \log \frac{v_{\mathscr{c}_k}}{v_\lambda} + \frac{d_x}{v_\lambda} \log \frac{v_{\mathscr{c}_k}}{v_\lambda} \\
        &+ \log \frac{v_{\mathscr{c}_k}}{v_{\mathscr{c}_k^\prime}} \sum_{x_i \in \mathcal{C}_k^\prime} \frac{d_{x_i}}{v_\lambda}  \\
        =& \frac{g_{\mathscr{c}_k^\prime}}{v_\lambda} \log \frac{v_{\mathscr{c}_k^\prime}}{v_\lambda}  - \frac{g_{\mathscr{c}_k}}{v_\lambda} \log \frac{v_{\mathscr{c}_k}}{v_\lambda} 
        + \frac{d_x}{v_\lambda} \log \frac{v_{\mathscr{c}_k}}{v_\lambda}
        + \frac{v_{\mathscr{c}_k^\prime}}{v_\lambda}  \log \frac{v_{\mathscr{c}_k}}{v_{\mathscr{c}_k^\prime}},\\
     \end{aligned}
\end{equation}
where, $d_x^\text{in}$ denotes the sum of edges from node $x$ to nodes $x_j \in \mathcal{C}_k$.
Since the $v_{\mathscr{c}_k}$, $g_{\mathscr{c}_k}$ and $d_x^\text{in}$ can be cached during the algorithm iterations:
\begin{equation}
       v_{\mathscr{c}_k^\prime} = v_{\mathscr{c}_k} - d_x, \quad
       g_{\mathscr{c}_k^\prime} = g_{\mathscr{c}_k} - 2* d_x^\text{in} + d_x. \\
\end{equation}

\begin{table}[t]
    \centering
    \caption{Glossary of Notations.}\label{tab:symbol}
    \label{TabNotations}
    \setlength{\tabcolsep}{1.6mm}
    \begin{tabular}{r|l}
    \toprule
   \textbf{Notation} & \textbf{Description} \\
    \midrule
    $G$, $G^d$ & Static network or graph, dynamic network\\
    $\mathcal{V}$ & Nodes (vertices) in network $G$\\
    $\mathcal{E}$ & Edges (links) in network $G$\\
    $\Delta \mathcal{V}$, $\Delta \mathcal{E}$ &  \makecell[l]{Variations of edges and nodes between consecutive\\ snapshots in dynamic network}\\
    $\mathcal{P}$ & A set of communities of network $G$\\
    $C_i$ & The $i$-th community in $\mathcal{P}$\\
    \midrule
    $S_i$ & The strategy of the node (player) $i$ \\
    $s$ & \makecell[l]{A strategy profile combines the strategies chosen \\by all the players in the game} \\
    $u_i$ & The payoff function of player $i$ \\
    $\varphi$ &  A potential function in the potential game \\
    \midrule
    $\mathcal{T}$ & An encoding tree of the graph $G$ \\
    $\alpha$ & A node in the encoding tree $\mathcal{T}$ \\
    $\bar{\alpha}$ & The parent node of the node $\alpha$ in the encoding tree \\
    $\lambda$ & The root node of the encoding tree $\mathcal{T}$ \\
    $T_\alpha$ & A set of vertices in the encoding tree node $\alpha$ \\
    $h(\alpha)$ & The height of the tree node $\alpha$ \\
    $\mathcal{H}(G)$ & \makecell[l]{The structural entropy (SE) of graph $G$ on \\the encoding tree} \\
    $\mathcal{H}^{(d)}(G)$ & The $d$-dimensional structural entropy of $G$ \\
    $\mathcal{H}^2(\mathcal{P})$ & The 2D structural entropy of the graph partition $\mathcal{P}$ \\
    $g_\alpha$ & \makecell[l]{The summation of the degrees (weights) of the cut \\edges of $T_\alpha$} \\
    $d_x$ & \makecell[l]{The degree of the node $x$ in the graph $G$} \\
    $v_\alpha, v_{\bar{\alpha}}, v_\lambda$ &  \makecell[l]{The summation of the degrees (weights) of encoding \\tree nodes, $\alpha$, $\bar{\alpha}$, and $\lambda$} \\
    $\Delta_{\text{S}}, \Delta_{\text{L}}, \Delta_{\text{T}}$ &  \makecell[l]{Heuristic functions for strategies: Stay, Leave and \\be alone, and Transfer to another community}\\
    $\Delta_{\text{O}}$ & Heuristic function for overlapping nodes \\
    $\tau_n$, $\tau_o$, $\gamma$, $r$ &  \makecell[l]{The termination threshold, the overlap threshold, \\the stable round threshold, and the  overlap factor} \\
    $\mathcal{V}^d$, $\mathcal{V}^{ind}$ & The directly and indirectly affected node set\\
    \bottomrule
\end{tabular}
\end{table}

\subsection{Pseudo-code of CIPA}
\label{addCIPA}

Algorithm~\ref{Alg:CIPA} presents the pseudo-code of the Cascading Influence Propagation-based Adaptive (CIPA) strategy. Given the community partition $\mathcal{P}{t-1}$ of the previous snapshot $G{t-1}$, CIPA first initializes the partition $\mathcal{P}_t$ for the current snapshot $G_t$ (Line~1), and then updates the per-community's volume and cut, according to edge/node additions and deletions (Lines~2–11).
In contrast to the static \framework, the CIPA-enabled dynamic \framework restricts the updates to the affected node set and iteratively refines it in an adaptive manner (Lines~12–32): it cascades the influence to identify additional potentially impacted nodes during the iterations, and removes nodes that have reached a stable state based on the stable-round threshold $r$. 
This preserves detection quality while substantially reducing redundant computations, thereby allowing the original algorithm to efficiently accommodate continuously evolving dynamic networks.

\begin{algorithm}[th]

    \SetAlgoShortEnd
    \caption{\textcolor{black}{CIPA for dynamic community detection.}\label{Alg:CIPA}}
    \let\oldnl\nl
    \newcommand{\nonl}{\renewcommand{\nl}{\let\nl\oldnl}}

    \DontPrintSemicolon
    \KwIn{Graph $G_t = (\mathcal{V}_t, \mathcal{E}_t)$ at time $t$;
        Changes in edges and nodes ($\mathcal{V}^{add}$, $\mathcal{E}^{add}$, $\mathcal{V}^{del}$, $\mathcal{E}^{del}$);
        Community partition $\mathcal{P}_{t-1}$ of $G_{t-1}$;
        Stable round threshold $r$.}
    \KwOut{community partition $\mathcal{P}_t$ of $G_t$.}

    Initialize  communities $\mathcal{P}_t \gets $ Eq. 16. \;
    \For{edge $e \in \mathcal{E}^{add} \cup \mathcal{E}^{del}$}{
        $x,y,w \gets e$ \Comment{Update volumes and cuts}\;
        \If{$e \in \mathcal{E}^{add}$}{
            $v_{c_x} = v_{c_x} +w$, $v_{c_y} = v_{c_y} +w$\;
            \If{$\mathcal{C}_x \neq \mathcal{C}_y$}{$g_{c_x} = g_{c_x} +w$, $g_{c_y} = g_{c_y} +w$ \;}
        }\Else{
            $v_{c_x} = v_{c_x} -w$, $v_{c_y} = v_{c_y} -w$\;
            \If{$\mathcal{C}_x \neq \mathcal{C}_y$}{$g_{c_x} = g_{c_x} -w$, $g_{c_y} = g_{c_y} -w$ \;}
        }
    }
    $\mathcal{V}^{d} \gets $ Eq. 15, $\mathcal{V}^{ind} \gets \{ \}$. \;
    Initialize state labels $ l \gets \{0 \mid x\in \mathcal{V}^{d}\}$.\;
    \While{true}{
        $M \gets 0$\;
        \For{node $x \in \mathcal{V}^d \cup \mathcal{V}^{ind}$}{
          Alg. 1, lines 7-22 \Comment{Potential Game} \;
            \If{$t \neq t_c$}{
                $M \gets M+1$, $l_x = 0$ \;
                
                \If{$x \notin  \mathcal{V}^{add}$}{
                    $\mathcal{V}^{ind} \gets$ Eq. 17\;
                } 
                \If{$x \in  \mathcal{V}^{ind}$}{
                    $\mathcal{V}^{ind} \gets \mathcal{V}^{ind} \setminus \{x\}$\;
                    $\mathcal{V}^{d} \gets \mathcal{V}^{d} \cup \{x\}$\;
                } 
            }
            \Else{
                $l_x \gets l_x + 1$\;
                \uIf{$x \in \mathcal{V}^{ind} \land l_x = 1$}{
                    $\mathcal{V}^{ind} \gets \mathcal{V}^{ind} \setminus \{x\}$
                }
                \uElseIf{$x \in \mathcal{V}^{d} \land l_x \geq r$}{
                    $\mathcal{V}^{d} \gets \mathcal{V}^{d} \setminus \{x\}$
                }
            }
        }
        
        \If{$M = 0$ or Eq. 13 is satisfied}{
            Break \;
        }
    }    
    \Return{$\mathcal{P}_t$}
    
\end{algorithm}

\subsection{Detailed Time Complexity Analysis}
\label{addTime}
\textbf{Static Community Detection. }
In the non-overlapping detection phase, the algorithm initially sets each node as an individual cluster and initializes community volumes and cut edge summations, which takes \(O(|\mathcal{V}|)\) time. 
For each node, the best strategy computation for each node depends on its degree, taking \(O(d_{\text{avg}})\) time, since we can compute \(\Delta_{\text{L}}(x, C)\) in \(O(1)\) time. 
Therefore, evaluating and updating all nodes in one iteration requires \(O(d_{\text{avg}} \cdot |\mathcal{V}| )\).
Overall, given that the algorithm runs for \(I\) iterations, the total time complexity is \(O(I \cdot d_{\text{avg}} \cdot |\mathcal{V}|)\). 
This complexity indicates that the algorithm's performance scales linearly with the number of nodes and their average degree, with the number of iterations needed for convergence influencing the overall computational effort. 
In the overlapping community detection phase, the algorithm’s time complexity is \(O(d_{\text{avg}} \cdot |\mathcal{V}| )\). 
This complexity arises because each node is processed by iterating over its adjacent communities.

\textbf{Dynamic Community Detection.}
In the dynamic community detection setting, the per-snapshot time complexity of our algorithm is reduced to
$O\!\left(I \cdot d_{\max} \cdot \left|\mathcal{V}^{d} \cup \mathcal{V}^{ind}\right|\right)$,
where $\mathcal{V}^{d}$ and $\mathcal{V}^{ind}$ denote the sets of directly and indirectly affected nodes in the current snapshot, respectively. $I$ is the maximum number of iterations, and $d_{\max}$ is the maximum node degree.
Specifically, initializing the community partition for the current snapshot takes
$O\!\left(\left|\mathcal{V}^{add}\right|\right)$ time.
Next, updating the community statistics (i.e., volumes and cuts) requires scanning the changed edges, which costs
$O\!\left(\left|\mathcal{E}^{add} \cup \mathcal{E}^{del}\right|\right)$,
where $\mathcal{E}^{add}$ and $\mathcal{E}^{del}$ denote the sets of added and deleted edges, respectively.
Moreover, since each indirectly affected node must be incident to at least one changed edge, we have
$\left|\mathcal{E}^{add} \cup \mathcal{E}^{del}\right| \leq \left|\mathcal{V}^{ind}\right|$,
and thus
$O\!\left(\left|\mathcal{E}^{add} \cup \mathcal{E}^{del}\right|\right)=O\!\left(\left|\mathcal{V}^{ind}\right|\right)$.
During the detection stage, the algorithm only processes nodes identified as affected, with  the cost of
$O\!\left(I \cdot d_{\max} \cdot \left|\mathcal{V}^{d} \cup \mathcal{V}^{ind}\right|\right)$.
Therefore, the total per-snapshot complexity is
\begin{equation}
\begin{aligned}
    &O\!\left(\left|\mathcal{V}^{add}\right| + \left|\mathcal{V}^{ind}\right| + I \cdot d_{\max} \cdot \left|\mathcal{V}^{d} \cup \mathcal{V}^{ind}\right|\right)\\
&= O\!\left(I \cdot d_{\max} \cdot \left|\mathcal{V}^{d} \cup \mathcal{V}^{ind}\right|\right),
\end{aligned}
\end{equation}
which indicates that the runtime is directly determined by the number of potentially affected nodes identified by the dynamic update mechanism.

\subsection{Datasets}
\label{add:Datasets}
We conduct comprehensive experiments on fourteen large-scale networks, which are described in detail as follows:

\begin{itemize}[left=0pt]
    \item \textbf{Amazon}~\cite{snapnets} is constructed by crawling the Amazon website, and each product category provided by Amazon defines a ground-truth community.
    
    \item \textbf{YouTube}~\cite{snapnets} is a video-sharing platform that also functions as a social network, allowing users to establish friendships and create groups that other users can join. 
    These user-defined groups are considered ground-truth communities.

    \item \textbf{DBLP}~\cite{snapnets} is a co-authorship network where two authors are connected if they have published at least one paper together. 
    Authors who have published in a specific journal or conference form a community.

    \item \textbf{LiveJournal}~\cite{snapnets} is a free online blogging community where users can declare friendships with each other. 
    User-defined friendship groups are considered ground-truth communities. 

    \item \textbf{Orkut}~\cite{snapnets} is a free online social network where users can form friendships and create groups that other members can join. 
    These user-defined groups are considered ground-truth communities. 
    \item \textbf{Friendster}~\cite{snapnets} is originally launched as a social networking platform, enabling users to establish friendships and create groups. The user-defined groups are regarded as ground-truth communities.
    
    \item \textbf{Wiki}~\cite{snapnets} is a directed overlapping network constructed from Wikipedia hyperlink data collected in September 2011. 
    This network comprises page titles and their associated category information, where each article may belong to multiple categories. 
    
    \item \textbf{X12\_static and X12}~\cite{mcminn2013building}.
     \textcolor{black}{X12\_static is a Twitter social network in which nodes correspond to user-generated tweets. 
     It comprises 68,841 English tweets collected over a 28-day period in 2012, encompassing 503 distinct event categories.
     In the X12\_static network, nodes are linked through shared attributes, such as common hashtags or user mentions.
     In addition, we employ SBERT~\cite{reimers2019sentence} to embed tweets, and compute pairwise cosine similarities. 
     For each node, the 150 most semantically similar nodes are selected as neighbors. 
     To construct the dynamic network X12, we transform X12\_static into a temporal network consisting of ten consecutive snapshots. 
     Each snapshot corresponds to a three-day interval, defined based on the posting timestamps of the tweets.}

    \item \textbf{X18\_static and X18}~\cite{mazoyer2020french}.
    \textcolor{black}{X18\_static is a Twitter social network consisting of 64,516 French tweets published over 23 days in 2018, covering 257 types of events.
    The construction method for the networks X18 and X18\_static is identical to that of X12 and X12\_static, with X18 comprising eight consecutive network snapshots.}

    \item \textbf{LFRs}. 
    \textcolor{black}{To simulate the properties of real-world complex networks, five groups of large-scale weighted synthetic networks are generated using the LFR benchmark tool\footnote{\url{https://github.com/eXascaleInfolab/LFR-Benchmark_UndirWeightOvp}}.  
    ~\cite{lancichinetti2009benchmarks}
    The networks contain 50K, 100K, 200K, 500K, and 1M nodes, respectively. 
    The mixing parameter $muw$ of the LFR benchmark tool is set to 0.6.}
    
\end{itemize}

\subsection{Baselines}
\label{add:baselines}
The detailed descriptions of all baselines are as follows:
\begin{itemize}[left=0pt]
    \item 
    \textbf{SLPA}~\cite{xie2011slpa} is an overlapping community detection method based on label propagation, designed to identify community structures in networks by propagating labels between nodes. 
    
    \item 
    \textbf{Bigclam}~\cite{yang2013overlapping} is an overlapping community detection method for large networks based on Non-negative Matrix Factorization (NMF), which maximizes the likelihood estimate of the graph to find the optimal community structure.

    \item 
    \textbf{NcGame}~\cite{ferdowsi2022detecting} is an overlapping community detection algorithm based on non-cooperative game theory, which treats nodes as self-interested participants to find communities that maximize individual node benefits.
    \item 
    \textbf{Fox}~\cite{lyu2020fox} is a heuristic overlapping community detection method that measures the closeness between nodes and communities by approximating the number of triangles formed by nodes and communities.

    \item 
    \textbf{Louvain}~\cite{blondel2008fast} is based on modularity optimization, which identifies non-overlapping community structures in networks through a multi-level optimization strategy.
    
    \item 
    \textbf{DER}~\cite{kozdoba2015community} is a diffusion entropy reduction graph clustering algorithm. 
    It employs random walks to embed the graph in a space of measures, after which a modified version of k-means is applied in that space.  

    \item 
    \textbf{Leiden}~\cite{traag2019louvain} is an improved version of the Louvain algorithm, which enhances community detection accuracy and stability through local optimization and refinement steps.
    \item 
    \textbf{FLPA}~\cite{traag2023large} is a fast variant of LPA \cite{Raghavan2007Near} that is based on processing a queue of nodes whose neighborhood recently changed. 

    \item 
    \textbf{QCA}~\cite{nguyen2011adaptive} is a rule-based modularity optimization algorithm that updates community structures according to predefined rules for node and edge additions and deletions.
    \item 
    \textbf{DynaMo}~\cite{zhuang2019dynamo} is a modularity-based method, which gradually maximizes modularity gain when updating the community structure of dynamic networks.

    \item 
    \textbf{DCDME}~\cite{sun2022dynamic} is based on the Matthew effect, which updates the community structure by iteratively processing all potentially affected communities in each snapshot.
    
    \item 
    \textbf{DCDBFE}~\cite{anuar2025bird} is inspired by the bird flock effect, leveraging natural ecological principles to identify and update communities in dynamic networks.
\end{itemize}

The open-source implementations provided by the CDlib library\footnote{\url{https://cdlib.readthedocs.io}} are used for SLPA and DER, while the Igraph library\footnote{\url{https://igraph.org/}} implementations are employed for Louvain, Leiden, and FLPA.
For BigClam, NcGame, Fox, QCA, DynaMo, DCDME, and DCDBFE, we used their respective open-source implementations\footnote{\url{https://github.com/snap-stanford/snap}}\textsuperscript{,}\footnote{\url{https://doi.org/10.1038/s41598-022-15095-9}}\textsuperscript{,}\footnote{\url{https://github.com/timgarrels/LazyFox}}\textsuperscript{,}\footnote{\url{https://github.com/nogrady/dynamo}}\textsuperscript{,}\footnote{\url{https://github.com/sunwww168/DCDME}}\textsuperscript{,}\footnote{\url{https://github.com/sitiharyanti/DCDBFE_2024}}.

\begin{figure*}[b]
\vspace{-4mm}
\centering
\subfloat[Ground truth.]{%
    \includegraphics[width=0.24\linewidth, trim={0 1cm 0 1cm}, clip]{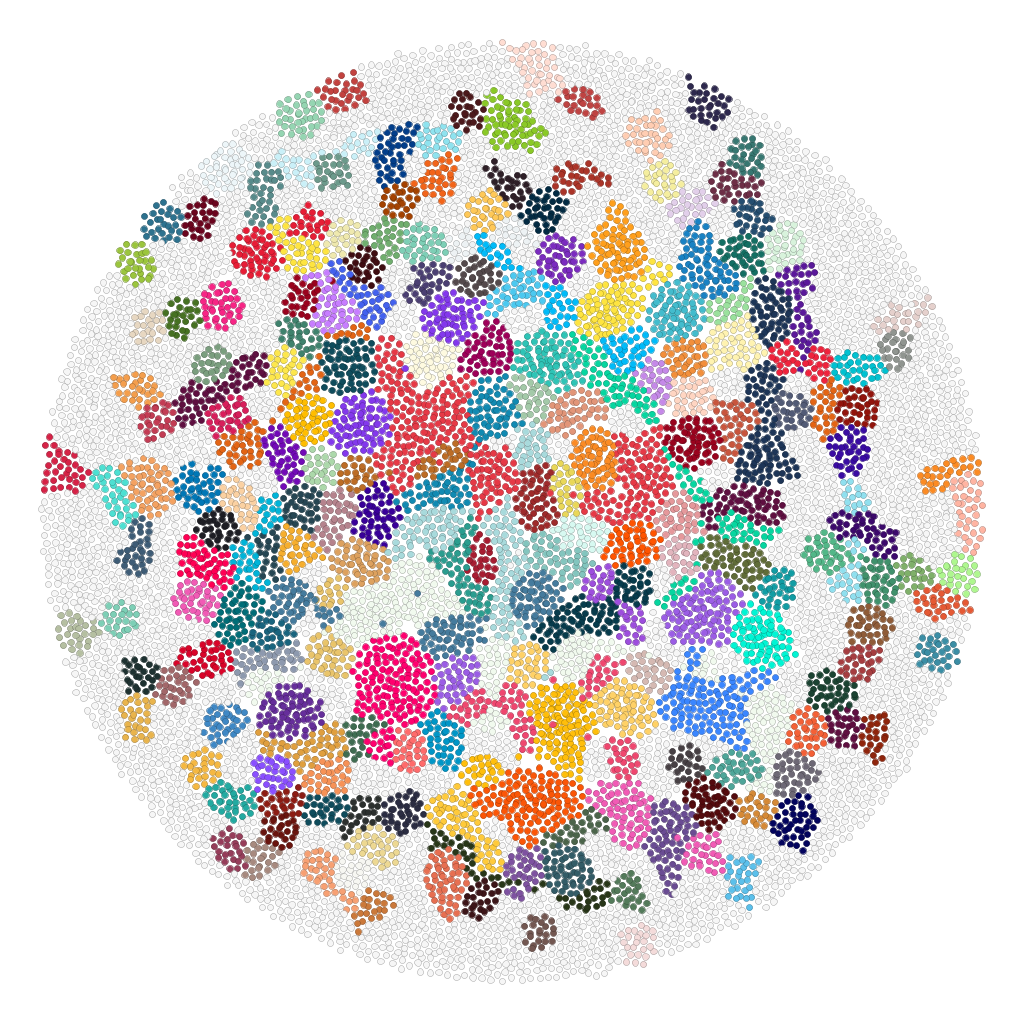}%
    \hspace{2mm}%
}
\subfloat[\framework .]{%
    \includegraphics[width=0.24\linewidth, trim={0 1cm 0 1cm}, clip]{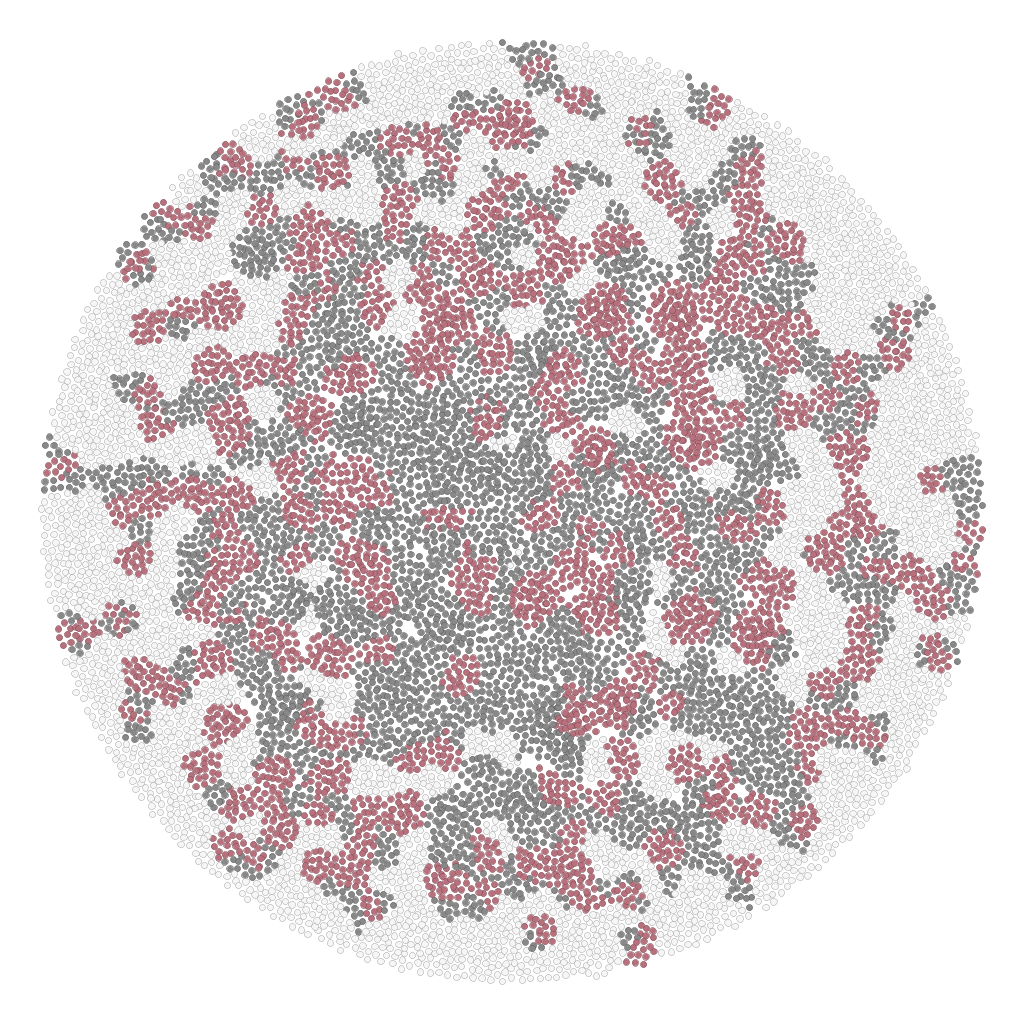}%
    \hspace{2mm}%
}
\subfloat[Fox.]{%
    \includegraphics[width=0.24\linewidth, trim={0 1cm 0 1cm}, clip]{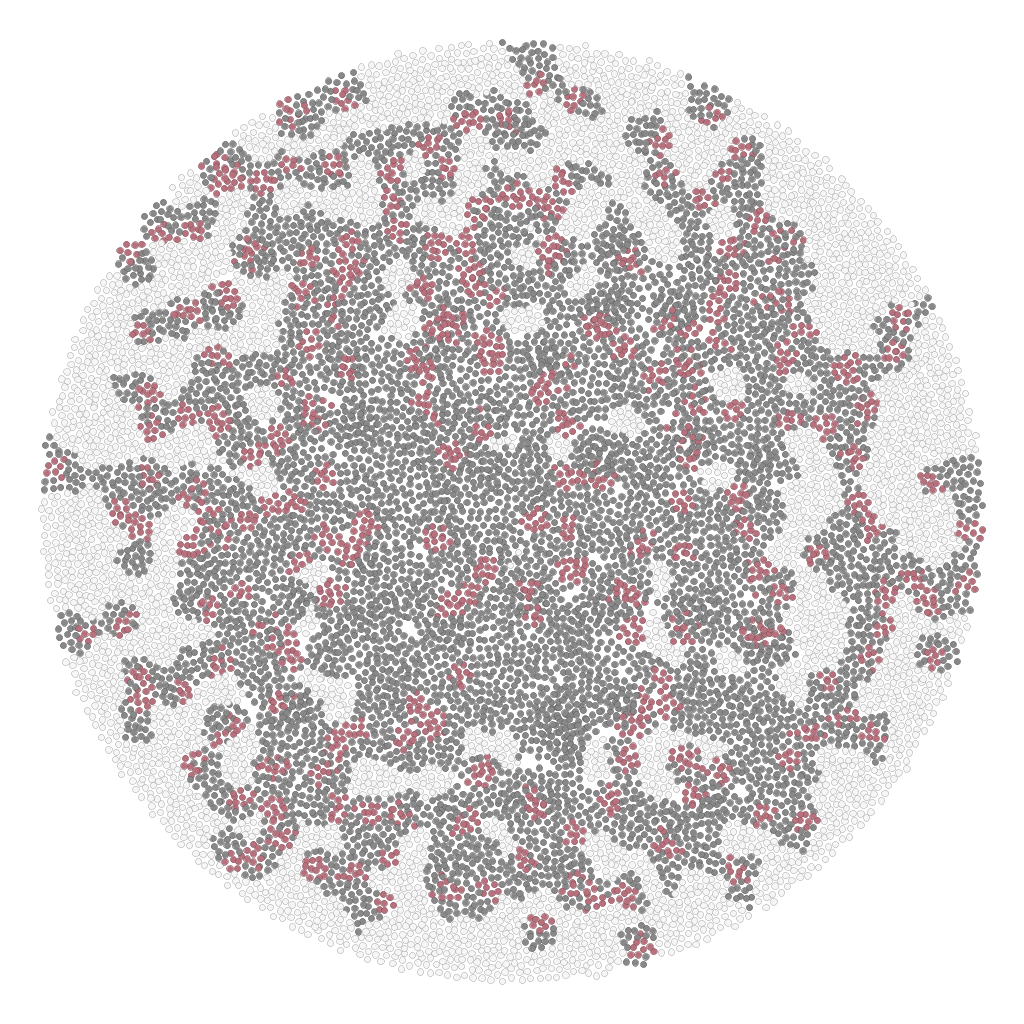}%
    \hspace{2mm}%
}
\subfloat[Bigclam.]{%
    \includegraphics[width=0.24\linewidth, trim={0 1cm 0 1cm}, clip]{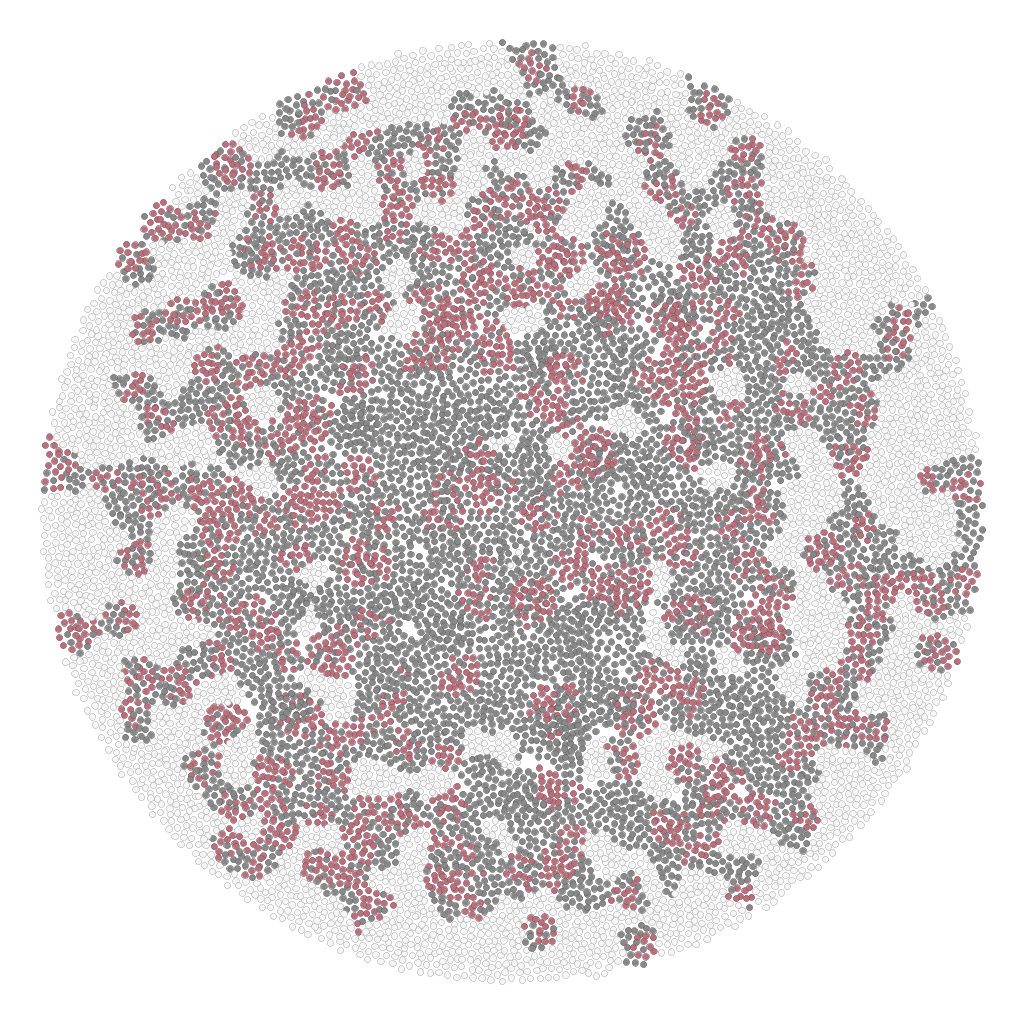}%
}
\vspace{-2mm}
\caption{ Community detection result visualization on Amazon. 
\colorbox[HTML]{b26c79}{Pink} represents nodes that are consistent with the ground truth communities, while \colorbox[HTML]{7f7f7f}{gray} denotes nodes that are inconsistent with the ground truth.\vspace{-4mm}}
\label{fig:Amazon_Top5000}
\end{figure*}

\begin{figure*}[b]
\centering
\subfloat[Ground truth.]{%
    \includegraphics[width=0.24\linewidth,  trim={0 1cm 0 1cm}, clip]{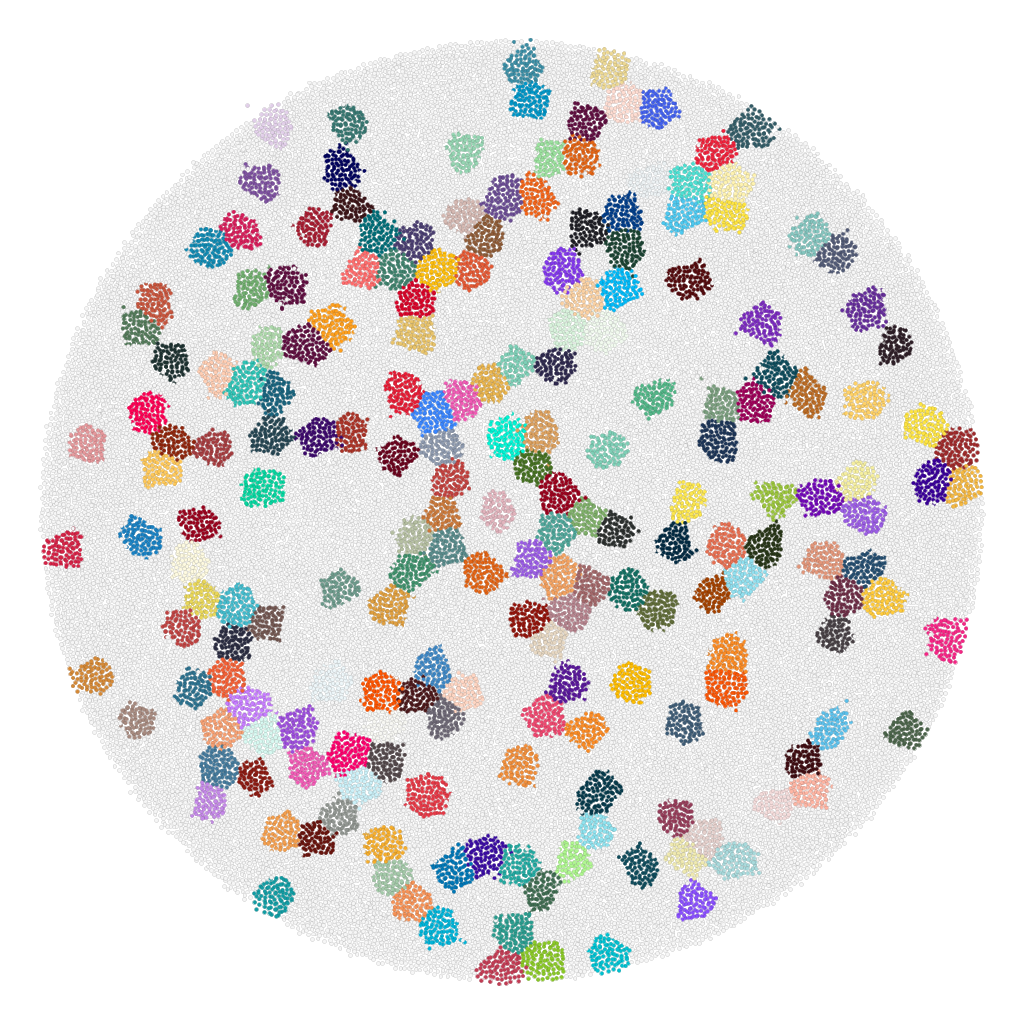}
    \hspace{2mm}%
}
\subfloat[\framework .]{%
    \includegraphics[width=0.24\linewidth,  trim={0 1cm 0 1cm}, clip]{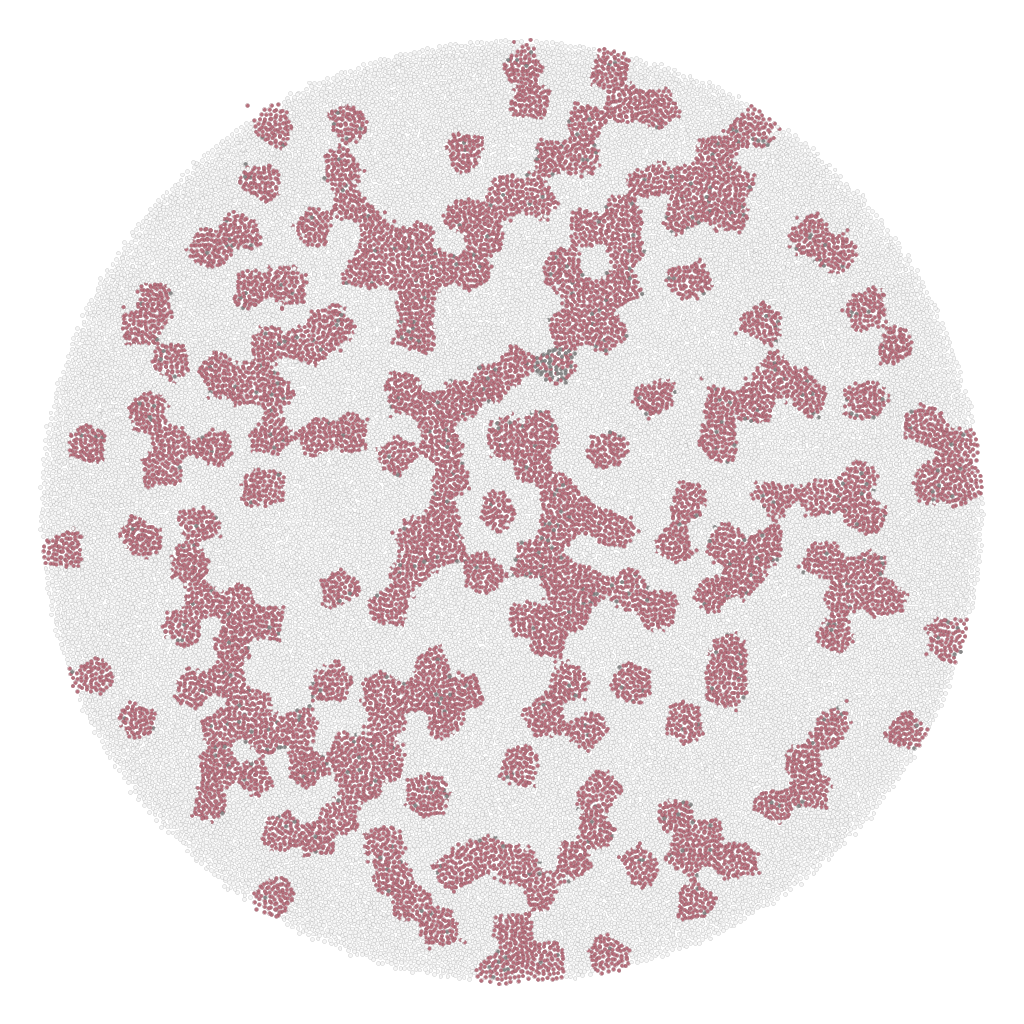}
    \hspace{2mm}%
}
\subfloat[FLPA.]{%
    \includegraphics[width=0.24\linewidth,  trim={0 1cm 0 1cm}, clip]{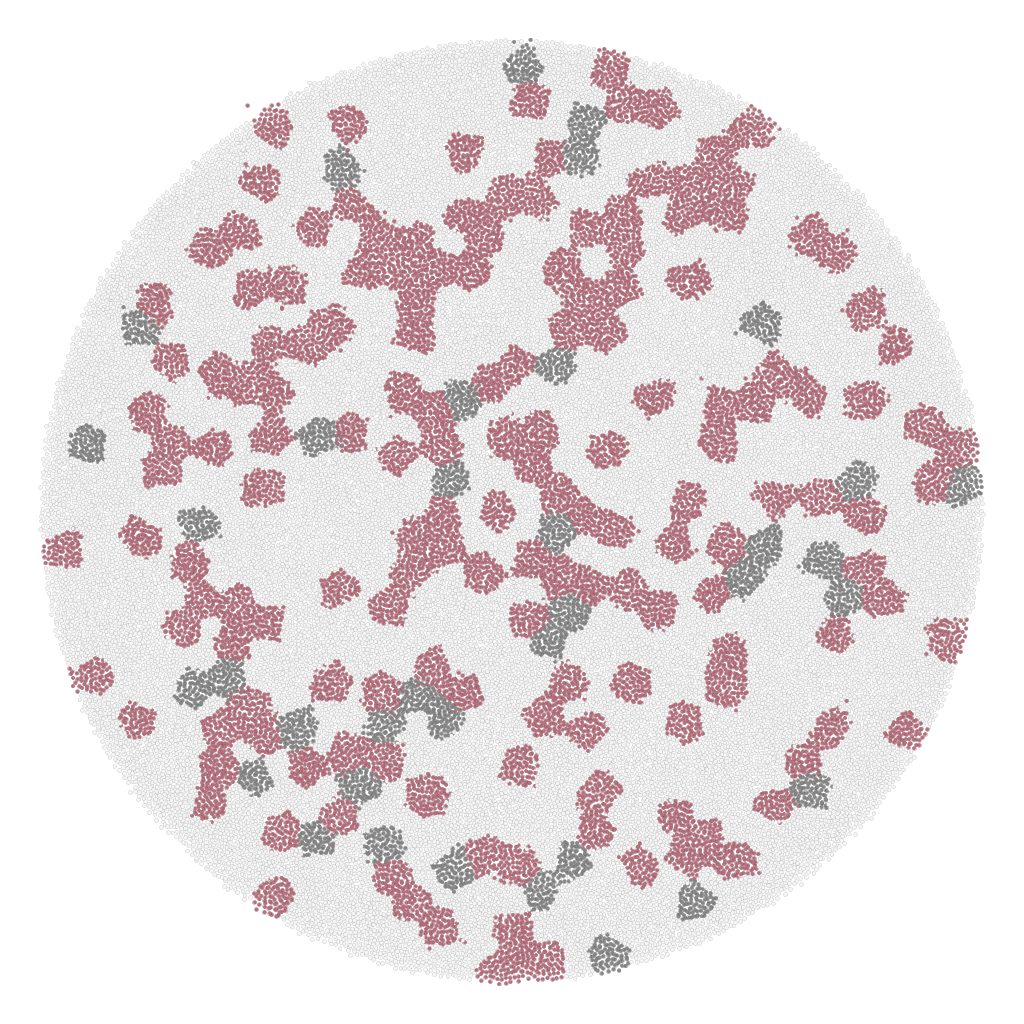}
    \hspace{2mm}%
}
\subfloat[DER.]{%
    \includegraphics[width=0.24\linewidth,  trim={0 1cm 0 1cm}, clip]{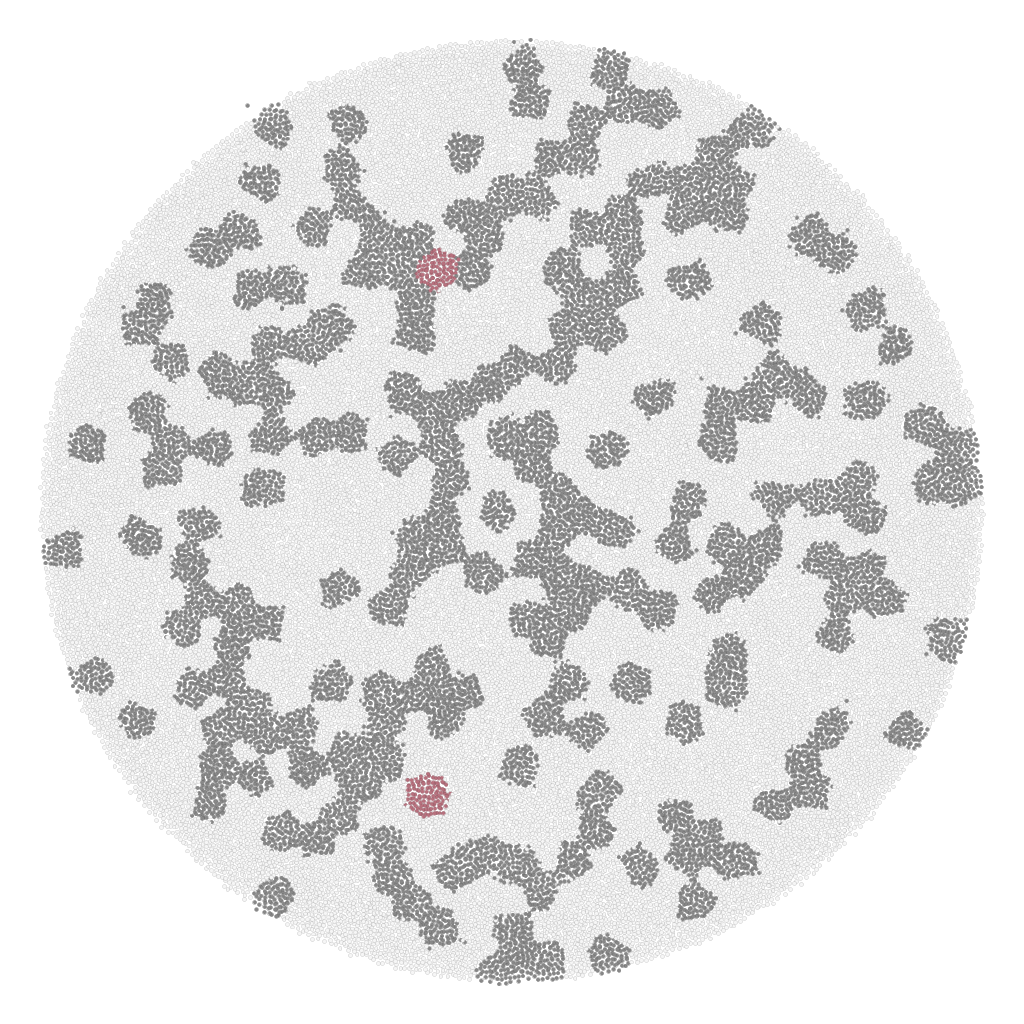}
    \hspace{2mm}%
}
\vspace{-2mm}
\caption{Community detection result visualization on LFR\_50K. 
\colorbox[HTML]{b26c79}{Pink} represents nodes that are consistent with the ground truth communities, while \colorbox[HTML]{7f7f7f}{gray} denotes nodes that are inconsistent with the ground truth.\vspace{-4mm}}
\label{fig:LFR}
\end{figure*}
\subsection{Details of Evaluation Metrics.}
\label{add:Evaluation Metrics}
Given a ground truth community \(\mathcal{C} \subseteq \mathcal{V}\) and a reconstructed community \(\mathcal{C}^\prime \subseteq \mathcal{V}\), the precision \(P(\mathcal{C}^\prime, \mathcal{C})\) and recall \(R(\mathcal{C}^\prime, \mathcal{C})\) are defined as follows:
\begin{equation}
    P(\mathcal{C}^\prime, \mathcal{C}) = \frac{|\mathcal{C} \cap \mathcal{C}^\prime|}{|\mathcal{C}^\prime|}, \quad
    R(\mathcal{C}^\prime, \mathcal{C}) = \frac{|\mathcal{C} \cap \mathcal{C}^\prime|}{|\mathcal{C}|}.
\end{equation}
Precision represents the fraction of the reconstruction in the ground truth, while recall represents the fraction of the ground truth in the reconstruction. These notions are often combined into a single number between 0 and 1, known as the \(F_1\)-score, defined as:
\begin{equation}
    F_1(\mathcal{C}^\prime, \mathcal{C}) = 2 \cdot \frac{P(\mathcal{C}^\prime, \mathcal{C}) \cdot R(\mathcal{C}^\prime, \mathcal{C})}{P(\mathcal{C}^\prime, \mathcal{C}) + R(\mathcal{C}^\prime, \mathcal{C})}.
\end{equation}
The \(F_1\)-score has the additional advantage of being symmetric, i.e., \(F_1(\mathcal{C}^\prime, \mathcal{C}) = F_1(\mathcal{C}, \mathcal{C}^\prime)\), and it equals $1$ if and only if the sets \(\mathcal{C}\) and \(\mathcal{C}^\prime\) are identical.
To evaluate the set of detected clusters, we define the \(F_1\) score for a collection of ground truth communities \(\mathcal{P}\) and a collection of detected communities \(\mathcal{P}^\prime\) \cite{yang2013overlapping,lutov2019accuracy}. The F1-score for detection is calculated as the average of two values: the F1-score of the best-matching ground-truth community for each detected community, and the F1-score of the best-matching detected community for each ground-truth community:
\begin{equation}
\begin{aligned}
    F_1(\mathcal{P}^\prime, \mathcal{P}) = \frac{1}{2} &\left( \frac{1}{|\mathcal{P}^\prime|} \sum_{\mathcal{C}^\prime \in \mathcal{P}^\prime} \max_{\mathcal{C} \in \mathcal{P}} F_1(\mathcal{C}^\prime, \mathcal{C}) \right. \\
    & \left. + \frac{1}{|\mathcal{P}|} \sum_{\mathcal{C} \in \mathcal{P}} \max_{\mathcal{C}^\prime \in \mathcal{P}^\prime} F_1(\mathcal{C}, \mathcal{C}^\prime) \right).
\end{aligned}
\end{equation}
We also use the ONMI based on information theory developed by Lancichinetti and Fortunato \cite{Lancichinetti2009Detecting} and later
refined by McDaid et al. \cite{mcdaid2013normalized}.
\begin{equation}
     \text{ONMI}(\mathcal{P}^\prime, \mathcal{P}) = \frac{I(\mathcal{P}^\prime, \mathcal{P})}{\max(H\left( \mathcal{P}^\prime), H(\mathcal{P}) \right)},
\end{equation}
where  \( H(\mathcal{P}) \) is the Shannon entropy of $\mathcal{P}$, and $I(\mathcal{P}^\prime, \mathcal{P}) $ is the mutual information.
It is important to note that the difference between NMI and ONMI lies in that nodes in \( \mathcal{P} \) and \( \mathcal{P}^\prime \) only belong to a single community.
For the specific calculation formula, refer to literature \cite{mcdaid2013normalized}. F1 measures the node-level detection performance, while NMI aims at the community-level detection performance. 

\subsection{Visualization Study}
\label{add:Visualization}
To intuitively assess the effectiveness of the proposed algorithm in overlapping community detection, we visualize the top 5000 real communities from the Amazon dataset, ranked by node count. 
The top 200 communities are highlighted in distinct colors, while the remaining nodes are displayed in light gray. Meanwhile, we visualize the detection results of \framework, the best-performing baseline (Fox), and the worst-performing baseline (Bigclam), showing their alignment with the real communities.
As depicted in Fig.~\ref{fig:Amazon_Top5000}, the detection results of \framework~exhibit the closest match to the real communities, followed by Bigclam. 
This demonstrates that the 2D SE effectively models communities in real-world networks, and that the heuristic structural entropy function, grounded in the latent game framework, enables efficient detection. 
Conversely, Fox performs the worst on this dataset, likely due to its reliance on local triangle counting for community membership determination, which fails in certain network structures. 
In contrast, \framework~shows greater robustness.
 
Similarly, \textcolor{black}{for the visualization of the LFR\_50K network, we compare the best-performing baseline, FLPA, with the worst-performing baseline, DER, among non-overlapping detection algorithms. 
As shown in Fig.~\ref{fig:LFR}, although FLPA achieves the best baseline performance, it still exhibits noticeable deviations from the ground-truth community structure. 
In contrast, the visualization results of \framework~align closely with the ground-truth community structure, showing only minimal node assignment discrepancies. 
Unlike synthetic networks, which typically have clearer structures, real-world networks tend to exhibit more intricate and irregular community configurations. Nevertheless, \framework~effectively captures the underlying community structure in both cases. 
Notably, DER performs poorly, clustering nearly all nodes into only two communities.
This limitation stems from its measure space embedding approach, which fails to capture subtle node distinctions in large-scale network structures. 
In conclusion, \framework~demonstrates superior robustness and accuracy in both overlapping and non-overlapping community detection tasks.}